\documentclass[12pt]{article}
\pdfoutput=1

\usepackage[english]{babel}
\usepackage[utf8]{inputenc} 
\usepackage{lmodern}
\usepackage{amsmath, amsthm, amssymb}
\usepackage{enumerate}
\usepackage{xcolor} % in the preamble
\definecolor{darkred}{RGB}{139,0,0} % define a dark red color
\usepackage[colorlinks=true, linkcolor=blue, citecolor=blue, urlcolor=darkred]{hyperref}
\usepackage{bookmark}
\usepackage{graphicx}
\usepackage{caption}
\usepackage{subfig}
\usepackage{cite}
\usepackage[title]{appendix}

\newcommand{\RomanNumeralCaps}[1]
    {\MakeUppercase{\romannumeral #1}}

\usepackage{amsfonts}
\usepackage{makeidx}
\usepackage{bbm}
\usepackage{slashed}
\usepackage{booktabs}
\usepackage{braket}
\usepackage{mathtools}
\usepackage{float}
\usepackage{diagbox}

\numberwithin{equation}{section}

\newcommand{\ev}{\;\text{eV}}

\newcommand{\beq}{\begin{equation}}
\newcommand{\eeq}{\end{equation}}

\begin{document}
\vspace{0.1cm}
\begin{center}
{\LARGE \bf  A simplest modular $S_3$ model for leptons}\\[2ex]
\vspace{1cm}
{\large D. Meloni\footnote{e-mail address: davide.meloni@uniroma3.it} and M. Parriciatu\footnote{e-mail address: m.parriciatu@studenti.unipi.it}} \\[5mm]

\textit{Dipartimento di Matematica e Fisica, 
Universit\`a di Roma Tre\\Via della Vasca Navale 84, 00146 Rome, Italy}\\
\vspace{16pt}

\begin{center}{\bf Abstract}
\begin{quote}
We present minimalist constructions for lepton masses and mixing based on flavour symmetry under the modular group $\Gamma_N$ of lowest level $N=2$. As opposed to the only existing model of $\Gamma_2\cong S_3$ formulated in a SUSY framework, the only non-SM field is the modulus $\tau$, and a generalised CP symmetry is implemented. Charged-leptons masses are reproduced through symmetry arguments, without requiring fine-tuning of the free parameters. As a result, all lepton observables (masses and mixing) are reproduced within $1\sigma$ experimental range using a minimum of nine free real parameters (including the real and imaginary parts of the modulus). A normal ordering for the neutrino masses is predicted. We also obtain predictions for the CP violating phases: the Dirac CP phase is predicted around $1.6\pi$, the Majorana phases lie in narrow regions near $\pm \pi$. The sum of neutrino masses is within the current bound at $\sim 0.09\ev$. Furthermore, we provide predictions for the neutrinoless double beta decay and tritium decay effective masses, around $20\,\text{meV}$. Given the reduced number of free input parameters as compared to the existing literature on modular $S_3$, this work renews interest for a unified predictive model of quark-lepton sectors based on $\Gamma_2\cong S_3$.
\end{quote}
\end{center}
\end{center}

%%%%%%%%%%%%%%%%%%%%%%%% 1.  INTRODUCTION   %%%%%%%%%%%%%%%%%%%%%%%%%%%%%%
%

\setcounter{footnote}{0}
\setcounter{tocdepth}{1}
{\let\clearpage\relax \tableofcontents}

% !TEX spellcheck = 
\section{Introduction}
In the recent past, a substantial effort went into the understanding of lepton mixing and masses through flavour symmetries. The main goal of model-builders is to explain at once both the charged-lepton mass hierarchy and the masses and mixing angles in the neutrino sector. 
The idea of employing flavour symmetries in Particle Physics can be traced back to the original  Froggatt-Nielsen mechanism \cite{Froggatt:1978nt}: the Standard Model (SM) is extended with the addition of new scalar fields (called “flavons”), generally sterile under the SM gauge group and charged under a $U(1)$ global symmetry. When the flavons get a vacuum expectation value (VEV), the $U(1)$ is broken and, after electroweak symmetry breaking, the Yukawa interactions give rise to the fermion mass matrices, with entries suppressed by the number of flavons employed in the $U(1)$ invariant Lagrangian. While the procedure is quite effective in modeling the hierarchies among fermion masses, the resulting mixing is generally small, thus disfavoring a correct description of the Pontecorvo-Maki-Nakagawa-Sakata $U_\text{PMNS}$ mixing matrix \cite{Maki:1962mu,Pontecorvo:1957qd}. In this respect, a suitable alternative has been provided by non-Abelian discrete symmetries \cite{Altarelli:2010gt}. 
These symmetries act linearly on the fields, which are supposed to belong to irreducible representations (irreps) of the group. After the flavons develop a VEV at some energy scale, their specific vacuum alignment in flavour space helps in shaping the mass matrices. Among the simplest discrete groups, the 
permutation groups like $S_3$ \cite{Feruglio:2007hi}, $A_4$ \cite{Altarelli:2005yx}, $S_4$ \cite{Lam:2008sh} etc. 
have been quite successful thanks to their ability in reproducing, at the leading order, approximate forms of the $U_\text{PMNS}$, which can then be made compatible with experiments through small perturbative corrections. Typical drawbacks of such approaches are related to the difficulty in reproducing the correct hierarchy among the lepton masses (for which an $U(1)$ or a $Z_N$ symmetry are usually invoked) and to the complicated scalar sector needed to correctly align the flavons in the flavour space.\newline
% \vspace{1mm}

In \cite{Feruglio:2017spp} a new promising direction to address the flavour problem was suggested. This “bottom-up" approach is based on modular invariance (which is an old idea in the context of String Theory \cite{Lauer:1989ax,Lauer:1990tm}): the Yukawa couplings of the SM become modular forms of level $N$, and functions of a complex scalar field $\tau$, (called {\it modulus}), which
acquires a VEV at some high-energy scale. They are supposed (together with matter fields) to  transform in irreps under the action of the finite modular group $\Gamma_N$.
To make the SUSY
action modular invariant, one requires the superpotential to be modular invariant, and the Kähler potential to be invariant up to a Kähler transformation. In some of its minimal realizations,
no flavons other than the $\tau$ are needed, and the VEV of the modulus
is the only source of flavour symmetry breaking. As opposed to models based on non-Abelian discrete symmetries, the Modular Symmetry is non-linearly realized. An interesting feature of $\Gamma_N$ is that, if we limit ourselves to study modular forms of level $N\le 5$, we benefit from the fact that the finite modular group $\Gamma_{N\le 5}$ is isomorphic to the non-Abelian discrete  groups $S_3$, $A_4$, $S_4$ and $A_5$ considered in past flavour models. Examples in this direction, for leptons and quarks, are given by models $\Gamma_2\cong S_3$ \cite{Kobayashi:2018vbk}, $\Gamma_3\cong A_4$ \cite{Feruglio:2017spp,Kobayashi:2018vbk,Kobayashi:2018wkl,Criado:2018thu,Kobayashi:2018scp,Okada:2018yrn,Okada:2019uoy,Ding:2019zxk,Kobayashi:2019xvz,Asaka:2019vev,Ding:2019gof,Zhang:2019ngf,King:2020qaj,Ding:2020yen,Asaka:2020tmo,Okada:2020brs,Yao:2020qyy,Okada:2021qdf,Nomura:2021yjb,Chen:2021prl,Nomura:2022boj,Gunji:2022xig}, $\Gamma_4\cong S_4$ \cite{Ding:2019gof,Penedo:2018nmg,Novichkov:2018ovf,deMedeirosVarzielas:2019cyj,Kobayashi:2019mna,King:2019vhv,Criado:2019tzk,Wang:2019ovr,Wang:2020dbp,Qu:2021jdy}, $\Gamma_5\cong A_5$ \cite{Criado:2019tzk,Novichkov:2018nkm,Ding:2019xna}. 
A very nice feature of
these models is that, in the limit of exact
supersymmetry, the superpotential structure is completely determined
by the modular symmetry, thus it does not receive further perturbative or non-perturbative corrections \cite{Feruglio:2019ybq}. A direct consequence is the
remarkably limited number of free parameters, hence these models are generally
more predictive than the models based on traditional non-Abelian discrete symmetries. Notice that, to further reduce the number of free parameters, one can additionally require the theory to be CP-symmetric, as was done in ref. \cite{Novichkov:2019sqv}.\footnote{The CP invariance in a modular context had also been studied in compactified String Theory, for some examples see refs. \cite{Baur:2019kwi} and \cite{Acharya:1995ag,Dent:2001cc,Giedt:2002ns}.} In this approach, the only source of CP-violation is the VEV of the modulus $\tau$, and the CP-symmetry forces the parameters of the theory to be real. In spite of its apparent suitability, some aspects of modular model building still remain problematic; among them, the possibility of reproducing the charged-leptons mass hierarchy without fine-tuning (but see \cite{Novichkov:2021evw} for remarkable results in this direction), the use of a single modulus $\tau$ for describing both quark and lepton sectors (see the recent work from ref. \cite{Benes:2022bbg}), the evaluation of Kähler and SUSY breaking corrections to the mass spectrum \cite{Chen:2019ewa,Criado:2018thu}.\newline
% \vspace{0.5mm}

In the present paper, we want to address the problem of building a modular lepton flavour model based on the $\Gamma_2\cong S_3$ group. 
To the knowledge of the present authors, this possibility has not been scrutinized in detail as for other groups and, for that, it deserves special attention. The only existing non-GUT\footnote{For modular $S_3$ GUT models see refs. \cite{Du:2020ylx,Kobayashi:2019rzp}.} model employing the $S_3$ modular group in a SUSY framework is the one from\footnote{Non-SUSY constructions that employ the modular $S_3$ group have been attempted in refs.\cite{Okada:2019xqk} and \cite{Mishra:2020gxg}.}ref.\cite{Kobayashi:2018vbk}. In this paper the authors consider three different combinations of the leptons assignments in the doublet and singlet,\footnote{In order to exclude degenerate eigenvalues for the charged leptons, the $SU(2)_L$ singlet fields are assigned to $S_3$ singlets, while the $SU(2)_L$ doublets are assigned to $S_3$ doublets or singlets.} while also introducing two additional flavons $\phi^{(1)},\phi^{(2)}$ besides the modulus $\tau$. 
The charged lepton matrix is diagonal while the neutrino mass matrix is obtained through the Weinberg operator. To get their results, the authors employed $13$ real parameters (including $\text{Re}\,\tau$ and $\text{Im}\,\tau$) and, as already mentioned, two ad-hoc flavons for the charged-lepton sector. The charged-leptons mass hierarchy is not accounted for by the symmetry, thus the parameters must be fine-tuned in order to reproduce the experimental values.\newline

Our purpose here is to go beyond the existing literature, proposing a modular model based on $S_3$ subject to the following conditions:
\begin{enumerate}
\item it employs the least number of free parameters (less than $12$, which is the number of lepton observables we wish to reproduce or predict: three charged-lepton masses, three neutrino masses, three angles and three CP violating  phases); 
\item it uses no flavons other than the modulus $\tau$;
\item it does not need extra matter fields;
\item it is less affected by fine-tuning among the free parameters.
\end{enumerate}

In this paper we show that this is indeed possible: we find two viable models for lepton masses and mixing with a maximum number of ten free real parameters and a minimum of nine. They are able to reproduce the charged-leptons mass hierarchy from symmetry arguments, with less severe fine-tuning of the free parameters, and the $U_\text{PMNS}$ mixing angles within 1 $\sigma$ experimental error. Both predict normal ordering for the neutrino masses, and a CP Dirac phase around $1.6\pi$ (which comes uniquely from the VEV of the modulus), largely compatible with current hints of maximal leptonic CP violation \cite{Capozzi:2021fjo}. We also have predictions for the two Majorana phases: both lie in quite narrow ranges near $\pm \pi$. The sum of neutrino masses, which is of cosmological interest, is predicted around $0.09\ev$ and is compatible with the recent upper bound of $0.115\,\text{eV}$ ($95\,\%\,\text{C.L.}$) from \cite{eBOSS:2020yzd}. The effective neutrino masses for neutrinoless double beta decay and tritium decay are both of order $\sim 20\,\text{meV}$, hence are largely compatible with present bounds \cite{KamLAND-Zen:2022tow}, \cite{KATRIN:2021uub}.\newline

Given the rather simple structure of the viable matrices we found, our constructions also allow for a more detailed analytical treatment of the diagonalisation, and, consequently, give a better understanding of the obtained numerical results.\newline
% \vspace{1mm}

The paper is organized as follows: in section \ref{2_frame} we review the features of modular symmetries and the group-theoretic aspects of $\Gamma_2\cong S_3$; in section \ref{4_strategy} we outline the strategies we used to construct the viable models; in sections \ref{5_CL} and \ref{6_neutr} we build the charged-leptons and neutrino mass matrices; in section \ref{7_num} we discuss the numerical scan approach and list the results. Finally, in \ref{concl} we draw our conclusions and discuss potential sources of corrections to our models. Two appendices are devoted to summarize the relevant features of the $S_3$ group.
%%%%%%%%%%%%%%%%%%%%%%%
%
\section{The Modular Framework in SUSY theories\label{2_frame}}
In the context of transformations of the modulus $\tau$, we conventionally call $\Gamma\equiv\text{SL}(2,\mathbb{Z})$ the \emph{homogeneous} modular group and $\overline\Gamma=\text{SL}(2,\mathbb{Z})/\{\pm\mathbbm{1}\}\equiv\text{PSL}(2,\mathbb{Z})$  the \emph{inhomogeneous} modular group. The \emph{inhomogeneous} modular group $\overline{\Gamma}$ is discrete, infinite and non-compact \cite{Feruglio:2019ybq}. We are interested in holomorphic functions which have the modulus $\tau$ as the argument, restricted to the upper-half of the complex plane:
\begin{equation}
\label{uppp}
\mathcal{H}=\{\tau\in\mathbb{C}\,|\, \text{Im}(\tau)>0\}\,.
\end{equation}
$\overline{\Gamma}$ acts on $\tau\in\mathcal{H}$ through the linear fractional transformation $\gamma\,:\, \tau\to \gamma(\tau)$ 

\begin{equation}
\label{gammatau}
\gamma(\tau)=\frac{a\tau+b}{c\tau+d}\quad,\quad a,b,c,d\in\mathbb{Z}\quad, \quad ad-bc=1\,,
\end{equation}
and it is generated by the following transformations:
\begin{equation}
\label{essti}
S\,:\,\tau \rightarrow -\frac{1}{\tau} \quad\quad,\quad\quad T\,:\,\tau\rightarrow \tau+1
\end{equation}
which, as can be verified, satisfy:
\begin{equation}
\label{relatt}
S^2=(ST)^3=\mathbbm{1}\,.
\end{equation}
These generators are represented in $\text{SL}(2,\mathbb{Z})$ by the $2\times2$ matrices
\begin{equation}
\label{esstimatrix}
S=\begin{pmatrix}
0 &1\\
-1&0
\end{pmatrix}\quad\quad,\quad\quad T=\begin{pmatrix}
1 &1\\
0&1
\end{pmatrix}\,.
\end{equation}
We now introduce the series of groups $\Gamma(N)$ for $N=1,2,3...$
\begin{equation}
\label{princN}
\Gamma(N)=\left\{\begin{pmatrix}
a& b\\
c&d
\end{pmatrix}\in \text{SL}(2,\mathbb{Z})\,\Big| \begin{pmatrix}
a& b\\
c&d
\end{pmatrix}\equiv\begin{pmatrix}
1& 0\\
0&1
\end{pmatrix}(\text{mod } N) \, \right\}\,,
\end{equation}
where the natural number $N$ is called the \emph{level}. It turns out that $\Gamma(1)=\Gamma$, but for $N\ge 2$ we have that $\Gamma(N)$ are infinite normal subgroups of $\Gamma\equiv \text{SL}(2,\mathbb{Z})$. The group $\Gamma(N)$ acts on $\tau\in\mathcal{H} $ through the linear fractional transformation \eqref{gammatau}, where $a,b,c,d$ satisfy $ad-bc=1$ with $a,d\equiv 1 (\text{mod } N)$ and $b,c\equiv 0 (\text{mod } N)$. We call \emph{modular form} a function $f(\tau)$ holomorphic in $\mathcal{H}$ and at $i\infty$ which, under $\gamma\in\Gamma(N)$ transforms as \cite{Gunning1962}:
\begin{equation}
\label{modforms}
f(\gamma(\tau))=(c\tau+d)^k f(\tau)\quad\quad,\quad \gamma=\begin{pmatrix}
a&b\\
c& d
\end{pmatrix}\,,
\end{equation}
where $k\ge 0$ is an integer called \emph{the weight}. We have no modular forms for $k<0$ \cite{Feruglio:2017spp}. From the definition follows that the product of two modular forms of the same level but with different weights $k$ and $k'$ is a modular form of weight $k+k'$. For $k=0$ only a constant satisfies the definition. Indeed, the modular property is extremely constraining and few functions satisfy it \cite{cohen:hal-01677348}. As discussed in \cite{Feruglio:2017spp, Gunning1962}, modular forms of weight $k$ and level $N$ form a linear space $\mathcal{M}_k(\Gamma(N))$ of finite dimension $d_{k}(\Gamma(N))$. In practice, the integer $d_{k}(\Gamma(N))$ specifies the number of linearly independent modular forms of level $N$ and weight $k$. Some examples are provided in table \ref{fertable}. We can also define $\overline{\Gamma}(N)\equiv \Gamma(N)/\{\pm \mathbbm{1}\}$ for $N=1$ and $N=2$, but since the element $-\mathbbm{1}$ does not belong to $\Gamma(N)$ for $N>2$, we identify $\Gamma(N)\equiv \overline{\Gamma}(N)$ for $N>2$. Now, from a model-building perspective, we may be interested in compact groups which admit unitary representations. Compact modular groups can be constructed by considering the quotient $\Gamma_N\equiv \overline{\Gamma}/\overline{\Gamma}(N)$ \cite{Feruglio:2017spp}: this is called the \emph{inhomogeneous} finite modular group and it admits finite-dimensional unitary representations. Its generators $S,T$ satisfy the conditions \eqref{relatt} plus the following: $T^N=\mathbbm{1}$ for $N\ge 2$. For $N\le 5$ the modular group $\Gamma_N$ is isomorphic to the familiar discrete permutation groups $S_3,A_4,S_4,A_5$, as summarised in table \ref{fertable}. We can also construct $\Gamma_N'\equiv {\Gamma}/{\Gamma}(N)$:  this is called the \emph{homogeneous} finite modular group. As discussed, for $N\le 2$ we have $\Gamma'_N=\Gamma_N$. For $N>2$ these groups are distinct. In particular $\Gamma'_N$ is the double-cover of $\Gamma_N$. More recently, the double-cover group $\Gamma_N'$ has been used in modular flavour models, see for example \cite{Liu:2019khw,Liu:2020akv,Wang:2020lxk,Yao:2020zml,Novichkov:2020eep,Novichkov:2021evw,deMedeirosVarzielas:2022fbw,Ding:2022nzn}. If we consider the group $\Gamma_N$ in model-building, the modular forms transform in irreps of this group only if the weight $k$ in \eqref{modforms} is an even number. However, if one considers the double cover group $\Gamma_N'$, the weights can be even or odd \cite{Liu:2019khw}.\newline

In our work we use the group $\Gamma_2\cong S_3$, which is its own double-cover. For $N=2$ we have that $-\mathbbm{1}\in \Gamma(N)$, hence a modular form of level $2$ transforms under $-\mathbbm{1}$ as
\begin{equation}
\label{transfodd}
f(\gamma(\tau))=(-1)^kf(\tau)
\end{equation}
But, since under $-\mathbbm{1}$ the modulus transforms  as:
$$\gamma(\tau)=\frac{-\tau}{-1}=\tau\quad\quad ,\quad \gamma=\begin{pmatrix}-1&0\\0&-1\end{pmatrix},$$
the transformation (\ref{transfodd}) implies
$$f(\tau)=(-1)^kf(\tau)$$
which, for $k$ odd, is satisfied only by $f(\tau)=0$. Hence, odd-weighted modular forms of level $2$ can only be zero. As a consequence, in our work the weights of the modular forms are constrained to be even numbers. \newline

Once a finite modular group has been chosen, one finds a basis of modular forms with the lowest weight that transform in the irreducible representations of the chosen discrete group. All the higher-weighted modular forms of the same level can be constructed as homogeneous polynomials of this basis. 

\subsection{The $\Gamma_2\cong S_3$ group \label{3_s3}}
The group $S_3$ is characterized by three irreducible representations: the doublet $\mathbf{2}$, the singlet $\mathbf{1}$ and the pseudo-singlet $\mathbf{1'}$ (a brief recap of its features is given in appendix \ref{appendix}). The lowest even-weighted modular forms are given by $k=2$ and, from table \ref{fertable}, we see that there are at most two independent modular forms which will constitute our basis. We will call them $Y_1(\tau)$ and $Y_2(\tau)$. In \cite{Feruglio:2017spp} a procedure to find the expression of these two modular forms was illustrated, making use of the Dedekind eta function which acts as a “seed" function for the modular forms of $\Gamma_2$. This seed function is defined in $\mathcal{H}$ as:
\begin{equation}
\label{eta}
\eta(\tau)\equiv q^{1/24}\prod_{n=1}^\infty(1-q^n)\quad\quad,\quad q\equiv e^{2\pi i\tau}
\end{equation}
and  under the generators $S$ and $T$ it transforms as: $\eta(-1/\tau)=\sqrt{-i\tau}\eta(\tau)$ and $\eta(\tau+1)=e^{i\pi/12}\eta(\tau)$, respectively. It turns out that the function $\eta^{24}(\tau)$ is a modular form of level $N=1$ and weight $12$. \newline
\begin{table}[]
\centering
\def\arraystretch{1.7}
\begin{tabular}{|c|c|c|}
\hline
$N$ & $d_k(\Gamma(N))$ & \multicolumn{1}{l|}{$\Gamma_N$} \\ \hline
$2$ & $k/2+1$          & $S_3$                           \\ \hline
$3$ & $k+1$            & $A_4$                           \\ \hline
$4$ & $2k+1$           & $S_4$                           \\ \hline
$5$ & $5k+1$           & $A_5$                           \\ \hline
\end{tabular}
\caption{\small{\it Properties of modular forms and of the finite modular groups of a given level $N\le 5$. Here $d_k(\Gamma(N))$ is the dimension of the linear space $\mathcal{M}_k(\Gamma(N))$, hence it also specifies the maximum number of independent modular forms of weight $k$. We assume $k$ is an even integer. In the last column we report the non-Abelian discrete groups isomorphic to $\Gamma_N$ for a given $N$.}}
\label{fertable}
\end{table}
\vspace{1mm}

In ref. \cite{Kobayashi:2018vbk} the lowest-weighted modular forms $Y_1(\tau)$ and $Y_2(\tau)$ have been calculated in the real-symmetric basis\footnote{For the isomorphism $\Gamma_2\cong S_3$, we shall choose the generators of $S_3$ such that their 2-dimensional representations satisfy the $\Gamma_2$ relations (\ref{relatt}) plus the $T^2=\mathbbm{1}_{2\times 2}$.} of $S_3$ reported in appendix \ref{appendix}. Their expressions are given in terms of the Dedekind's eta:\footnote{In \eqref{multiplets1} $\eta'$ denotes the first derivative of $\eta$ with respect to $\tau$.}
\begin{equation}
\label{multiplets1}
\begin{cases}
Y_1(\tau)=\displaystyle\frac{c}{2}\left[\frac{\eta'(\tau/2)}{\eta(\tau/2)}+\frac{\eta'\left(\frac{\tau+1}{2}\right)}{\eta\left(\frac{\tau+1}{2}\right)}-\frac{8\eta'(2\tau)}{\eta(2\tau)}\right]\\ \\
Y_2(\tau)=\displaystyle\frac{c}{2}\sqrt{3}\left[\frac{\eta'(\tau/2)}{\eta(\tau/2)}-\frac{\eta'\left(\frac{\tau+1}{2}\right)}{\eta\left(\frac{\tau+1}{2}\right)}\right]
\end{cases}\,,
\end{equation}
where $c$ is an arbitrary normalisation constant not fixed in the procedure used to calculate the expressions for $Y_1(\tau)$, $Y_2(\tau)$. To the knowledge of the present authors there is no prescription for “$c$" from a bottom-up approach. In \cite{Feruglio:2017spp} that was fixed to $c=i/(2\pi)$, while a more detailed discussion on the normalisation was given in ref.\cite{Novichkov:2019sqv} in the context of the generalised CP-symmetric (gCP) modular theories; however the arbitrariness still remains. 
It can be noticed that the choice does not modify the internal structure of the mass matrices if the singlet contractions in flavour space happen between irreducible representations that include all three lepton generations, since in that case “$c$" can be factored out as an overall coefficient. This is easy to achieve with discrete groups that are equipped with the triplet $\mathbf{3}$. However this is not the case with $S_3$, in which one needs at least two distinct irreps for the fermion fields, hence the choice of “$c$" can potentially play a non-trivial role in model-building with this group. In that regard, our particular choice is motivated in section \ref{4_strategy}.\newline
\vspace{1mm}

The two forms in \eqref{multiplets1} transform as an $S_3$ doublet, expressed with the short-hand notation:\footnote{In the object $Y_{\boldsymbol\rho}^{(a)}(\tau)$, the symbol $\boldsymbol{\rho}$ labels the irreducible representation, while the upper index inside parentheses $(a)$ stands for the number of modular forms of weight 2 involved in the tensor-product construction of $Y_{\boldsymbol\rho}^{(a)}(\tau)$. Hence, the weight is $2a$. On the other hand, the individual components are denoted by $Y_{j}^p(\tau)$, where $j=\{1,2\}$, and $p$ is simply the exponent.} 
\begin{equation}
\label{ydoublets}
Y^{(1)}_\mathbf{2}(\tau)\equiv\begin{pmatrix}Y_1(\tau)\\
Y_2(\tau)
\end{pmatrix}_\mathbf{2}\,.
\end{equation}
According to the $S_3$ tensor decompositions (reported in appendix \ref{cgc}), all the higher-weighted modular forms transforming in the representations of $\mathbf{1}$ and $\mathbf{1'}$ can be obtained from the doublet in \eqref{ydoublets}.

\subsection{$S_3$ Modular Forms of Higher Weights \label{highers3}}
As discussed, the doublet \eqref{ydoublets} has modular weight $k=2$ and is the generator multiplet of the ring $\mathcal{M}(\Gamma(2))$, i.e. we can construct all the higher-weighted modular multiplets of level $2$ from the tensor products of \eqref{ydoublets}. Given that the linear space $\mathcal{M}_k(\Gamma(2))$ has dimension $k/2+1$ (with $k$ even, see table \ref{fertable}) we can construct at most $3$ linearly independent modular forms of weight $4$ in the following way:
\begin{equation}
\label{weight2}
Y^{(1)}_\mathbf{2}(\tau)\otimes 
Y^{(1)}_\mathbf{2}(\tau)=(Y_1^2(\tau)+Y_2^2(\tau))_\mathbf{1}\oplus \begin{pmatrix}
Y_2^2(\tau)-Y_1^2(\tau)\\
2Y_1(\tau)Y_2(\tau)
\end{pmatrix}_\mathbf{2}\,,
\end{equation}
in which the pseudo-singlet combination $\mathbf{1'}$ vanishes identically due to the antisymmetry of its indices. In what follows, we will express them with the already mentioned notation:
\begin{equation}
Y^{(2)}_\mathbf{2}(\tau)\equiv  \begin{pmatrix}
Y_2^2(\tau)-Y_1^2(\tau)\\
2Y_1(\tau)Y_2(\tau)
\end{pmatrix}_\mathbf{2}\quad\quad,\quad\quad Y^{(2)}_\mathbf{1}(\tau)\equiv (Y_1^2(\tau)+Y_2^2(\tau))_\mathbf{1}\,.
\end{equation}
Likewise, we expect $4$ linearly independent modular forms of weight $6$, and we can obtain them by combining three forms of weight $2$. A straightforward calculation yields:
\begin{equation}
\begin{split}
\label{sixweight}
Y^{(2)}_\mathbf{2}(\tau)\otimes Y^{(1)}_\mathbf{2}(\tau)=[3Y_1(\tau)Y_2^2(\tau)-Y_1^3(\tau)]_\mathbf{1}\oplus [Y_2^3(\tau)-3Y_2(\tau)Y_1^2(\tau)]_\mathbf{1'}\oplus\\ \oplus \begin{pmatrix}
Y_1(\tau)(Y_1^2(\tau)+Y_2^2(\tau))\\
Y_2(\tau)(Y_1^2(\tau)+Y_2^2(\tau))
\end{pmatrix}_\mathbf{2}\,,
\end{split}
\end{equation}
which will be denoted as $Y^{(3)}_\mathbf{1}$, $ Y^{(3)}_\mathbf{1'}$, and $Y^{(3)}_\mathbf{2}$, respectively. As the number of independent modular forms of weight $k$ grows with $k/2+1$, we obtain more ways to construct multiplets that transform in a given irreducible representation. As a consequence, there are more independent singlet contractions in the superpotential, and these come with additional free parameters. Hence, for the sake of minimality, we choose to employ a maximum weight of $6$.

\subsection{The SUSY Action}
We now turn to the modular model-building in a SUSY framework. In general one assumes that chiral superfields (which contain the SM fields) transform under the modular group $\Gamma_N$, and that the Yukawa couplings $Y(\tau)$ are vector-valued modular forms: they are holomorphic functions of the modulus $\tau$, which in turn is considered a chiral supermultiplet\footnote{It is assumed that the modulus $\tau$ is a dimensionless chiral supermultiplet, function of the superspace coordinates $x^i,\theta^\alpha,\bar\theta_{\dot\alpha}$. We can always introduce a fundamental scale $\Lambda_\tau$ and define a new supermultiplet as $\sigma=\Lambda_\tau\,\tau$, having now mass-dimension of one in natural units \cite{Feruglio:2017spp}.} whose scalar component is restricted to $\mathcal{H}$. The modular symmetry spontaneously breaks when the modulus acquires a VEV $(\braket{\tau}=\text{Re}\,\tau+i\text{Im}\,\tau)$ at some high-energy scale.\footnote{The dynamical description of the modular symmetry-breaking and the value of the high-energy scale where this symmetry breaks are currently unknown. In modular model-building, the VEV of the modulus is considered a free parameter allowed to vary to search the agreement with experimental data. However, there is some ongoing research on the scalar potential responsible for this VEV, for examples see refs. \cite{Kobayashi:2019xvz}, and \cite{Novichkov:2022wvg,Knapp-Perez:2023nty,Ishiguro:2020tmo,Ishiguro:2022pde}.} The most general supersymmetric action (turning off gauge interactions) that is also modular-invariant, in the context of $\mathcal{N}=1$ SUSY, can be written as:
\begin{equation}
\label{action}
\mathcal{S}=\int d^4x\int d^2\theta d^2\bar\theta \,K(\Phi,\bar\Phi)+\left[\int d^4x\int d^2\theta\, \mathcal{W}(\Phi)+\text{h.c.}\right]\,,
\end{equation}
where $\theta,\bar\theta$ are the Grassmann spinor-coordinates of the superspace; $\Phi=(\tau,\varphi)$ denotes the chiral superfields of the theory, and $\varphi$ is a shorthand for all the usual matter supermultiplets; $K(\Phi,\bar\Phi)$ is the Kähler potential that gives rise to the kinetic terms of the fields; $\mathcal{W}(\Phi)$ is the superpotential. Under the modular group $\bar\Gamma$, the chiral superfields transform as
\begin{equation}
\label{moddd_transf}
\begin{cases}
\tau\to \gamma(\tau)=\displaystyle\frac{a\tau+b}{c\tau+d}\\ \\
\varphi^{(I)}\to (c\tau+d)^{-k_I}\rho^{(I)}(\gamma)\varphi^{(I)}
\end{cases},\quad\text{with}\quad \gamma=\begin{pmatrix}a & b \\ c & d\end{pmatrix} \in \bar\Gamma\,,
\end{equation}
where the letter $I$ indicates a specific chiral superfield sector (e.g. the $SU(2)_L$ lepton doublets and singlets, Higgs doublets and so forth). The symbol $\rho^{(I)}(\gamma)$ stands for a unitary irreducible representation of $\Gamma_N$. A crucial feature of the modular symmetry framework is that the chiral superfields $\varphi^{(I)}$ are not modular forms, even though they transform very similarly in (\ref{moddd_transf}). Thus, the values of the numbers $(-k_I)$ (which will be called \emph{modular charges} to distinguish them from modular weights) have no a priori restriction, i.e. they can also be negative \cite{Feruglio:2017spp}. A particular model is completely specified once the superfields are assigned to the irreps and the modular charges are chosen. The action is invariant under the transformation (\ref{moddd_transf}) if the superpotential $\mathcal{W}(\Phi)$ is invariant and the Kähler potential acquires a shift:
\begin{equation}
\label{ccaler}
\begin{cases}
\mathcal{W}(\Phi)\to\mathcal{W}(\Phi)\\ \\
K(\Phi,\bar\Phi)\to K(\Phi,\bar\Phi)+f(\Phi)+\bar{f}(\bar\Phi)
\end{cases}\,.
\end{equation}
where $f(\Phi)$ is a holomorphic function of the superfields. The superpotential $\mathcal{W}(\Phi)$ will in general be a gauge-invariant combination of Yukawa modular forms $Y_{I_1...I_n}(\tau)$ and chiral superfields
$$\mathcal{W}(\Phi)=\sum (Y_{I_1...I_n}(\tau)\,\varphi^{(I_1)}...\varphi^{(I_n)})_\mathbf{1}\,,$$
where it is intended that the sum is performed over all combinations of superfields $\{I_1,...,I_n\}$ contracted with the $Y_{I_1...I_n}(\tau)$ to form all the independent singlets $\mathbf{1}$ of $\Gamma_N$. To counterbalance the transformations \label{mod_transf} of the chiral superfields, the Yukawa modular forms $Y_{I_1...I_n}(\tau)$ of weight $k_Y$ and level $N$ transform under the modular group as:
\begin{equation}
\label{yuk_transf}
Y_{I_1...I_n}(\tau)\to (c\tau+d)^{k_Y}\rho(\gamma)Y_{I_1...I_n}(\tau)\,,
\end{equation}
such that the superpotential is invariant if and only if the following conditions are met:
\begin{equation}
\label{vincoli}
\begin{cases}
\rho\otimes \rho_{I_1}\otimes \rho_{I_2}...\otimes \rho_{I_n} \supset\mathbf{1}\\
k_Y=k_{I_1}+k_{I_2}+...+k_{I_n}
\end{cases}\,,
\end{equation}
i.e. the modular weights of the Yukawa couplings counterbalance the modular charges of the chiral superfields, and there exists a $\Gamma_N$-invariant singlet contraction among all the tensor products. This requirement alone heavily constrains the superpotential: although there is an infinity of possible choices for the modular charges and field representations, the superpotential can only have a finite number of possible contributions (see discussion in ref. \cite{Liu:2021gwa}). The number of allowed operators in $\mathcal{W}(\Phi)$ is sharply restricted since the modular forms $Y(\tau)$ belong to the linear space $\mathcal{M}_{k}(\Gamma(N))$ of finite dimension $d_k(\Gamma(N))$. Aiming at a minimalistic model, the form of the modular-invariant Kähler potential is usually chosen to be the simplest possible (even though this is only a choice):
\begin{equation}
\label{kahler}
K(\Phi,\bar\Phi)=-h\Lambda_\tau^2 \log(-i\tau+i\bar\tau)+\sum_I(-i\tau+i\bar\tau)^{-k_I}{|\varphi^{(I)}|^2}\,,
\end{equation}
where $h$ is a positive constant and $\Lambda_\tau$ has the dimension of a mass.

\section{Our Building Strategy \label{4_strategy}}
Since the charged-leptons observables are known with great precision, we studied several field representations and modular charge assignments that led to the correct mass hierarchy among them, $m_e/m_\mu\approx 1/200$, $m_\mu/m_\tau\approx 1/17$, and based our strategy on the requirement that the charged lepton mass hierarchy has to be reproduced with a small amount of fine-tuning among the model free parameters, and that the smallest number of free parameters should be employed to reach our aim.\newline
\vspace{1mm} 

With these principles in mind we were able to single out the most suitable superfield assignments, summarised in table \ref{irrepss}.
\begin{table}[t]
\aboverulesep = 0pt
\belowrulesep = 0pt
\centering
\def\arraystretch{1.3}
\resizebox{0.8\textwidth}{!}{
\begin{tabular}{@{}|l|c|c|c|c|c|c|@{}}
\toprule
\multicolumn{1}{|c|}{} &
  \multicolumn{1}{c|}{$E_1^c$} &
  \multicolumn{1}{c|}{$E_2^c$} &
  \multicolumn{1}{c|}{$E_3^c$} &
  \multicolumn{1}{c|}{$D_\ell$} &
  \multicolumn{1}{c|}{$\ell_3$} &
  \multicolumn{1}{c|}{$H_{d,u}$} 
\\ \midrule
$\scriptstyle{SU(2)_L\times U(1)_Y}$ &  $(\mathbf{1},+1)$       & $(\mathbf{1},+1)$       & $(\mathbf{1},+1)$        & $(\mathbf{2},-1/2)$    & $(\mathbf{2},-1/2)$ &   $(\mathbf{2},\mp1/2)$   \\ \midrule
$\Gamma_2\cong S_3$ &  $\mathbf{1}$     & $\mathbf{1'}$       & $\mathbf{1'}$       & $\mathbf{2}$        & $\mathbf{1'}$        & $\mathbf{1}$  \\ \midrule 
$k_I$ &  $k_{E_1}$       & $k_{E_2}$       & $k_{E_3}$       & $k_D$        & $k_\ell$        & $k_d,k_u$   \\  \bottomrule 
\end{tabular}}
\caption{\small{\it Chiral supermultiplets, transformation properties under $\Gamma_2\cong S_3$ and modular charges.}}
\label{irrepss}
\end{table}
The superfields $E_i^c$ correspond to the three flavours of right-handed charged leptons (respectively $\{i=1,2,3\}\equiv \{e,\mu,\tau\}$), with (unspecified) modular charges $k_{E_1},k_{E_2},k_{E_3}$. The left-handed $SU(2)_\text{L}$ doublets
$$\ell_i=\begin{pmatrix}
\nu_i\\ E_i
\end{pmatrix}$$
are grouped into a doublet and a pseudo-singlet of $S_3$:
$$D_\ell\equiv \begin{pmatrix}
\ell_1 \\ \ell_2
\end{pmatrix}\sim \mathbf{2}\quad\quad,\quad\quad \ell_3\sim \mathbf{1'}\,,$$
with modular charges respectively $k_D$ and $k_\ell$. The Higgs doublets are completely sterile with respect to the modular symmetry: they transform as invariant singlets and from now on we will set $k_u=k_d=0$. 
The superpotential will be the sum of the charged-leptons and neutrino sectors. In our model, neutrino masses arise from the Weinberg operator \cite{Weinberg:1979sa}, so in general we will write (symbolically):
\begin{equation}
\label{general}
\mathcal{W}=\sum_i\alpha_i(E^c_iH_d L f_E(Y))_\mathbf{1}+\frac{g_i}{\Lambda}(H_uH_uLL f_W(Y))_\mathbf{1}\,,
\end{equation}
where $L$ is a placeholder for the left-handed multiplets $D_\ell$ or $\ell_3$, and $f_E(Y)$, $f_W(Y)$ stand for certain combinations of the modular multiplets $Y_\mathbf{\rho}^{(a)}(\tau)$, and $\alpha_i,g_i$ are free parameters. $\Lambda$ accounts for the smallness of the neutrino mass scale and is probably related to the violation of lepton number. In (\ref{general}), all the constraints for the modular charges from eq.(\ref{vincoli}) and the singlet contractions between the irreps of $S_3$ are to be understood in order to have a modular-invariant superpotential. The sum is performed over all possible singlet contractions from the assignments in table \ref{irrepss}. In an attempt to declutter the notation, we hid the necessary contractions for gauge and Lorentz invariance, but these are to be understood henceforth. After electroweak symmetry breaking, we obtain the mass lagrangian for the leptons: 
$$\mathcal{L}\supset -(M_\ell)_{ij}\bar{\ell_R}_i {\ell_L}_j-\frac{1}{2}(m_\nu)_{ij}\overline{\nu_L^c}_i{\nu_L}_j+\text{h.c.}\,,$$
written in the Weyl spinor notation of the SM. The  obtained  mass matrices are:
\begin{equation}
\label{masses}
(M_\ell)_{ij} = v_d\,y^e_{ij} \qquad (m_\nu)_{ij} = v_u^2\,y^\nu_{ij}/\Lambda\,,
\end{equation}
where $v_{d,u}$ are the VEV's of the $H_{d,u}$ Higgs doublets and $y^e_{ij},y^\nu_{ij}$ are Yukawa matrices 
depending on the VEV of the modulus $\tau$. \newline
\vspace{1mm}

In our construction, we implemented a gCP-symmetry. Given the $S_3$ basis used in this paper (reported in appendix \ref{appendix}), a CP transformation on the lowest-weight modular multiplets that is consistent with modular invariance \cite{Novichkov:2019sqv} yields:
\begin{equation}
\label{cp_vio}
Y(\tau)\xrightarrow{\text{CP}} Y(-\tau^*)=Y^*(\tau)\,.
\end{equation}
In our case, this relation is satisfied if we normalise the components in (\ref{multiplets1}) by choosing “$c$" purely imaginary.\footnote{Given the reality of the Clebsch-Gordan coefficients (shown in appendix \ref{cgc}), the transformation property \eqref{cp_vio} is then satisfied for all modular multiplets of higher weight \cite{Novichkov:2019sqv}.} With this in mind, the theory is CP-invariant if the action (\ref{action}) is sent to itself under a CP-transformation of  the modular forms as in eq.(\ref{cp_vio}) and of the superfields $\tau,\varphi_i$. As shown in \cite{Novichkov:2019sqv}, with our choices this is the case only if the superpotential couplings are real. Hence, the number of free real parameters (besides the modulus) is automatically reduced. As a consequence, $\text{Re}\,\tau$ is the only source of CP-violation. As it will become clear in the following, the request of having $\mathcal{O}(1)$ coupling ratios in the charged-lepton sector can be accommodated by choosing the following normalisation for the multiplets in (\ref{multiplets1}):
\begin{equation}
\label{cinorm}
c=i\frac{7}{25\pi}\,.
\end{equation}
As a consequence, the lowest-weight doublet has the following q-expansion:
\begin{equation}
\label{qexp}
\begin{pmatrix}
Y_1(\tau)\\ Y_2(\tau)
\end{pmatrix}_\mathbf{2}=\begin{pmatrix}
\displaystyle\frac{7}{100}+\frac{42}{25}q+\frac{42}{25}q^2+\frac{168}{25}q^3+...\\ \\
\displaystyle\frac{14\sqrt{3}}{25}q^{1/2}(1+4q+6q^2+...)
\end{pmatrix}\,.
\end{equation}
where $q\equiv \exp(2\pi i \tau)$. As discussed in ref.\cite{Novichkov:2018ovf}, the range of the modulus $\tau$ can be restricted to what is called the {\it fundamental domain}, denoted as $\mathcal{D}$:
\begin{equation}
\label{fun_domain}
\mathcal{D}=\left\{\tau\in \mathbb{C}\,:\, \text{Im}\,\tau>0\,,\, \,\,\left|\text{Re}\,\tau\right|\le \frac{1}{2}\,,\,\,\,|\tau|\ge 1 \right\}\,.
\end{equation}
In fact, thanks to modular symmetry, every non-zero VEV of the modulus $\tau\in\mathcal{H}$ can be mapped to $\tau'\in \mathcal{D}$. \newline
\vspace{1mm}

\begin{figure}
  \begin{center}
    \includegraphics[scale=0.1]{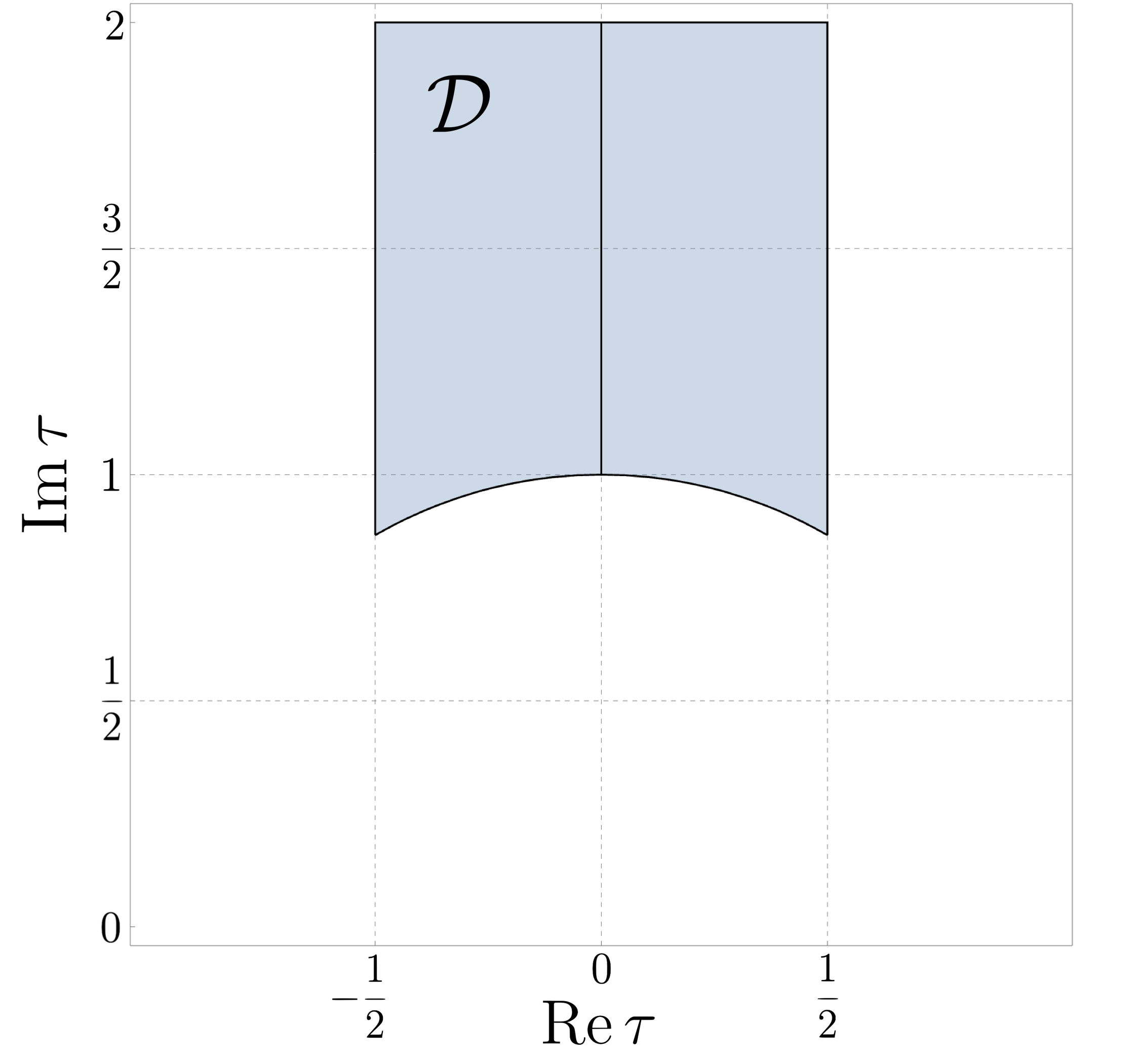}
    \caption{\small{\it The fundamental domain of the modulus $\tau$ as defined in eq.(\ref{fun_domain}).}}
    \label{ffund}
  \end{center}
  \end{figure}
Since the q-expansions are exponentially suppressed by $\text{Im}\,\tau$, the quantity $\epsilon\equiv Y_2(\tau)/Y_1(\tau)$, where $Y_2(\tau)$ and $Y_1(\tau)$ are the components in \eqref{qexp}, is always suppressed for $\tau\in \mathcal{D}$. Hence, $|\epsilon|$ can be used as an expansion parameter in our constructions. For instance, for $\text{Im}\,\tau \gtrsim 1.3$ we have $|\epsilon| \lesssim \lambda$, where $\lambda\approx 0.22$ is the Cabibbo angle, a condition that is independent from the choice of the normalization \eqref{cinorm} for the modular multiplets.

\section{Charged-leptons Sector \label{5_CL}}

The choice \eqref{cinorm} for $c$ is particularly suitable as we intend to use the modular weights in a way similar to the Froggatt-Nielsen charges, as will be clear later. In that regard, it should be noted that similar attempts were made in refs. \cite{Criado:2019tzk,King:2020qaj,Kuranaga:2021ujd} but, as opposed to those studies, we will not introduce “weightons" or additional flavons to drive the hierarchy. The hierarchy could instead be encoded in the fact that, for the VEV of $\tau$ inside the fundamental domain \eqref{fun_domain}, the $q$ in the expansions \eqref{qexp} becomes exponentially suppressed and we obtain $|Y_2|\lesssim |Y_1|$. 
Notice that a large number of successful models in the literature lie very close to the point $\tau=i$, as was recently pointed out by Feruglio et al. in refs. \cite{Feruglio:2021dte,Feruglio:2023bav}. 
To maintain the number of independent parameters at a reasonable level and to naturally obtain hierarchical eigenvalues in the charged lepton sector, we adopt the following strategy:
\begin{itemize}
\item we impose zeros in appropriate matrix entries through modular-symmetry constraints;
\item some matrix entries are exponentially suppressed with respect to other entries for $\tau\in\mathcal{D}$. Given the normalisation choice (\ref{cinorm}), this suppression will be more severe for entries which involve modular components that appear with larger exponents.\footnote{Those exponents, as can be seen in \eqref{sixweight}, will depend on the choices of the modular weights, thus they are sensible to the fields representations assignments and can be fixed by requesting modular invariance in the superpotential.}
\end{itemize}

 In the charged-leptons sector and before imposing any restrictions on the weights, the most general superpotential compatible with modular invariance and the fields assignments in table \ref{irrepss} is given by:
 \begin{multline}
 \label{general_CL}
 \mathcal{W}_e^{H}=\alpha E_1^cH_d(D_\ell Y_\mathbf{2}^{(a)})_\mathbf{1}+\beta E_2^cH_d(D_\ell Y_\mathbf{2}^{(b)})_\mathbf{1'}+\gamma E_3^cH_d\ell_3 Y_\mathbf{1}^{(c)}+\\ +\alpha_D E_1^cH_d\ell_3 Y_\mathbf{1'}^{(d_1)}+\beta_D E_2^c H_d\ell_3 Y_\mathbf{1}^{(d_2)}+\gamma_D E_3^cH_d(D_\ell Y_{\mathbf{2}}^{(d_3)})_\mathbf{1'}\,,
 \end{multline}
 where the letters inside the parentheses $(n)$ denote “$n$ insertions" of modular forms of weight $2$ (for a total weight of $2n$), and $\alpha,\beta,\gamma...$ are free parameters which, as discussed in the previous section, are taken real. As displayed in \eqref{vincoli}, modular invariance requires each operator in the superpotential to be weightless. Given the field assignments reported in table \ref{irrepss} and setting $k_D=k_\ell$ for the modular charges of the $SU(2)$ doublets $D_\ell, \ell_3$, we found two promising configurations of modular charges for the right-handed fields $E_1^c,E_2^c,E_3^c$ which allow us to implement the strategy discussed above: we will refer to these models as {\it Hierarchical} (${\cal O}(1)$ ratios among the Yukawa parameters) and {\it Minimal} (minimum set of free parameters).

\subsection{Charged-leptons: Hierarchical model \label{hierac_mod}}
If we wish to reproduce the hierarchical mass pattern while also maintaining ${\cal O}(1)$ Yukawa couplings ratios and a minimal number of free parameters, we can allow the modular charges of the right-handed fields to vary with the constraints:
\begin{equation}
\label{consmodr}
\begin{cases}
k_{E_1}=6-k_\ell\\
k_{E_2}=2-k_\ell\\
k_{E_3}=-k_\ell
\end{cases}\,.
\end{equation}
As a consequence, the invariant superpotential is
\begin{equation}
\label{ncl}
\mathcal{W}_e^{H}=\alpha E_1^cH_d(D_\ell Y_\mathbf{2}^{(3)})_\mathbf{1}+\beta E_2^cH_d(D_\ell Y_\mathbf{2})_\mathbf{1'}+\gamma E_3^cH_d\ell_3+\alpha_D E_1^cH_d\ell_3 Y_\mathbf{1'}^{(3)}\,.
\end{equation}
Notice that, compared to the general \eqref{general_CL}, the prohibition of the $\gamma_D$ term comes from the fact that, since there is no modular doublet of weight zero, there is no way to obtain a $S_3$ singlet from the tensorial contraction of $E_3^c$ ($\mathbf{1'}$) with $D_\ell$ ($\mathbf{2}$); in addition, the prohibition of the $\beta_D$ term comes from the fact that a modular singlet of weight $2$ does not exist: the first singlet has weight $4$, as seen in section \ref{highers3}. The resulting mass matrix after electroweak symmetry breaking is (written in right-left notation):
\begin{equation}
\label{mass_ncl}
M_\ell=\begin{pmatrix}
\alpha (Y^{(3)}_\mathbf{2})_1 & \alpha(Y^{(3)}_\mathbf{2})_2 & \alpha_D Y^{(3)}_\mathbf{1'} \\ \beta Y_2&-\beta Y_1& 0 \\ 0 & 0 & \gamma
\end{pmatrix}v_d\,,
\end{equation}
where the modular forms of weight $6$ in the first row are given by eq. \eqref{sixweight}:
$$\begin{pmatrix}
(Y^{(3)}_\mathbf{2})_1\\ (Y^{(3)}_\mathbf{2})_2
\end{pmatrix}_\mathbf{2}=\begin{pmatrix}
Y_1(Y_1^2+Y_2^2)\\
Y_2(Y_1^2+Y_2^2)
\end{pmatrix}_\mathbf{2}\quad,\quad Y_\mathbf{1'}^{(3)}=(Y_2^3-3Y_1^2Y_2)_\mathbf{1'}\,.
$$
The matrix \eqref{mass_ncl} can be expressed as follows:
\begin{equation}
\label{mass_par}
M_\ell=v_d\alpha Y_1^3\begin{pmatrix}
(1+\epsilon^2) & \epsilon(1+\epsilon^2) & -A\epsilon(3-\epsilon^2) \\ B \epsilon&-B& 0 \\ 0 & 0 & C
\end{pmatrix}\,,
\end{equation}
where $A\equiv \alpha_D/\alpha$, $B\equiv (\beta/\alpha)Y_1^{-2}$, $C\equiv (\gamma/\alpha)Y_1^{-3}$ and $\epsilon$ is the expansion parameter defined in section \ref{4_strategy}. Assuming $\frac{\beta}{\alpha}\sim\frac{\gamma}{\alpha}\sim\frac{\alpha_D}{\alpha}\approx 1$, we have $A\sim \mathcal{O}(1)$ and, given the $q$-expansions \eqref{qexp}, we also have $|B|,|C|\gg 1$ for $\tau\in\mathcal{D}$. With these orders of magnitude in mind, the approximated eigenvalues can be expressed in powers of $\epsilon$ as follows:
\begin{align}
\label{eigenhierarc}
&m_e=v_d\alpha \left(|Y_1^3|+\frac{3}{2}|Y_1^3||\epsilon|^2+\mathcal{O}(\epsilon^3)\right)\\
&m_\mu=v_d\alpha \left(|Y_1|+\frac{1}{2}|Y_1| |\epsilon|^2+\mathcal{O}(\epsilon^3)\right)\\
&m_\tau=v_d\alpha \left(1+\frac{9A^2}{2}|Y_1^6||\epsilon|^2+\mathcal{O}(\epsilon^3)\right)
\end{align}
and the hierarchical pattern $(m_\tau,m_\mu,m_e)\sim m_\tau(1,|Y_1|,|Y_1|^3)$ naturally arises with $|Y_1|\approx 7/100$. The matrix structure \eqref{mass_ncl} gives a non-trivial contribution to the PMNS matrix defined as $U_\text{PMNS}=U_\ell^\dagger U_\nu\text{diag}(e^{i\alpha_1},e^{i\alpha_2},1)$, where $\alpha_1,\alpha_2$ are the Majorana phases. Since we have $A/|C|^2\ll |\epsilon|$, the unitary matrix $U_\ell$ that diagonalises $M_\ell^\dagger M_\ell$ can be expressed as follows:
\begin{equation}
\label{unitarymin}
U_\ell\sim\begin{pmatrix}
1-\displaystyle\frac{|\epsilon|^2}{2}& -\overline{\epsilon} &0\\
\epsilon & 1-\displaystyle\frac{|\epsilon|^2}{2} &0\\
 0& 0& 1
\end{pmatrix}+\mathcal{O}(\epsilon^3)\,,
\end{equation}
up to $\mathcal{O}(1)$ multiplicative coefficients in front of $\epsilon,|\epsilon|^2$ terms. Then, we see that the charged lepton mixing matrix would provide a negligible contribution to the neutrino mixing but for the solar angle, for which we expect a small $\mathcal{O}(\lambda)$ contribution for $\text{Im}\,\tau$ sufficiently large (as in our previous example, $\text{Im}\tau \gtrsim 1.3$ is enough). Hence, we expect the leading contribution for the mixing angles to come from the neutrino sector through $U_\nu$. As will become clear in the following, since $k_{\ell}$ controls the models in the neutrino sector, the modular charges (\ref{consmodr}) will be completely determined once we select a specific neutrino model that produces the correct phenomenology.

\subsection{Charged-leptons: Minimal model}
If we choose to rely on a minimalistic construction, we may want to prohibit the $\alpha_D$ term in \eqref{ncl} by setting:
\begin{equation}
\label{minvinc}
\begin{cases}
k_{E_1}=4-k_{\ell}\\
k_{E_2}=2-k_{\ell}\\
k_{E_3}=-k_\ell
\end{cases}\,,
\end{equation}
and this works since a $\mathbf{1'}$ modular form of weight $4$ does not exist. We found that, among several possible $S_3$ realizations, this model is equipped with the smallest number of free parameters. Hence we call it \emph{Minimal}. The superpotential now reads:
\begin{equation}
\label{mcl}
\mathcal{W}_e^{M}=\alpha E_1^cH_d(D_\ell Y_\mathbf{2}^{(2)})_\mathbf{1}+\beta E_2^cH_d(D_\ell Y_\mathbf{2})_\mathbf{1'}+\gamma E_3^cH_d\ell_3\,,
\end{equation}
and the mass matrix is given by:
\begin{equation}
\label{mass_mcl}
M_\ell=\begin{pmatrix}
\alpha (Y^{(2)}_\mathbf{2})_1 & \alpha(Y^{(2)}_\mathbf{2})_2 &0 \\ \beta Y_2&-\beta Y_1& 0 \\ 0 & 0 & \gamma
\end{pmatrix}v_d\,.
\end{equation}
Following the same arguments from the previous section, and assuming $\mathcal{O}(1)$ for the coupling ratios, we find a mass pattern $(m_\tau,m_\mu,m_e) \sim m_\tau(1,|Y_1|,|Y_1|^2)$. Hence, the observed hierarchy can be reproduced by requesting $\beta/\alpha\sim \gamma/\alpha \approx \mathcal{O}(10)$ for the free parameters:
\begin{align}
\label{minimaleigenfull}
\centering
&m_e=v_d\alpha \left(|Y_1^2|-\frac{7}{2}|Y_1^2||\epsilon|^2+\mathcal{O}(\epsilon^3)\right)\\
&m_\mu=v_d\alpha \left(\frac{\beta}{\alpha}|Y_1|+\frac{1}{2}\frac{\beta}{\alpha}|Y_1| |\epsilon|^2+\mathcal{O}(\epsilon^3)\right)\\
&m_\tau=v_d\alpha\left(\frac{\gamma}{\alpha}\right)\,.
\end{align}
The mass matrix \eqref{mass_mcl} is also diagonalised by \eqref{unitarymin}. As mentioned in the previous section, the modular charges of the right-handed fields \eqref{minvinc} will be completely fixed by a choice for $k_\ell$ once a viable model for neutrino masses is found.\newline
\vspace{1mm}

\section{Neutrino Sector \label{6_neutr}}
The most general Weinberg superpotential that we assume to be responsible for neutrino masses is given by:\footnote{Notice that the combination $(D_\ell D_\ell)_\mathbf{1'}$ vanishes identically due to the antisymmetry of the $\mathbf{1'}$ tensorial contractions (see appendix \ref{cgc}).}
\begin{multline}
\label{general_wein}
\mathcal{W}_\nu\supset \frac{g}{\Lambda}H_uH_u(D_\ell D_\ell)_\mathbf{2} Y^{(w)}_\mathbf{2}+\frac{g'}{\Lambda}H_uH_uD_\ell \ell_3 (Y^{(q)}_\mathbf{2})+\\+ \frac{g''}{\Lambda}H_uH_u(D_\ell D_\ell)_\mathbf{1} Y^{(p)}_\mathbf{1}+\frac{g_p}{\Lambda}H_uH_u\ell_3\ell_3 (Y^{(d_w)}_\mathbf{1})\,,
\end{multline}
where, as usual, $\Lambda$ is the scale of new physics, presumably close to the GUT scale. The existence of the four terms in $\mathcal{W}_\nu$ implies constraints on the modular charges of the form: $k_D=w,~k_D+k_\ell=2q,~k_\ell=d_w,~k_D=p$; recalling that $k_D=k_\ell$, we end up with the relation:
\begin{equation}
\label{wein_const}
w=d_w=q=p=k_\ell\,.
\end{equation}
In order to avoid trivial structures in the Weinberg operator, we assume $k_\ell\ge 0$. Then, by varying $k_\ell$, we obtain all the possibilities for the neutrino masses and mixing:
\begin{itemize}
\item \textbf{W0} $(k_\ell=0)$: With this configuration the $g'$ term in (\ref{general_wein}) is forbidden because there are no doublet modular forms at weight $0$. Similarly, the term in $g$ is forbidden. Thus, we get:
$$\mathcal{W}^{k_\ell=0}_\nu\supset g''\frac{1}{\Lambda}H_uH_u(D_\ell D_\ell)_\mathbf{1}+{g_p}\frac{1}{\Lambda}H_uH_u\ell_3\ell_3\,,$$
which, after electroweak symmetry breaking, results in the following mass matrix:
$$m_\nu^{k_\ell=0}=\frac{2g''v_u^2}{\Lambda}
\begin{pmatrix}
1 & 0 &0\\
0 & 1 &0 \\
0 &0 &\frac{g_p}{g''}
\end{pmatrix}\,.$$
This model should be discarded since it predicts two exactly degenerate neutrino masses, contradicting the data from oscillation experiments. 
\item  \textbf{W1} $(k_\ell=1)$: Now it is the $g_p$ term that is forbidden since a singlet modular form of weight $2$ does not exist.\footnote{Notice that we could not simply use $\ell_3\ell_3\sim \mathbf{1}$ since the modular charges do not sum up to zero given that $k_\ell=1$. A modular form is needed to counterbalance the charges.} For the same reason $g''$ is also forbidden. Hence, we have:
$$\mathcal{W}^{k_\ell=1}_\nu\supset g\frac{1}{\Lambda}H_uH_u(D_\ell D_\ell)_\mathbf{2} Y_\mathbf{2}+{g'}\frac{1}{\Lambda}H_uH_uD_\ell \ell_3 Y_\mathbf{2}\,,$$
where the $g'$ term is the singlet contraction arising from $\mathbf{2}\otimes \mathbf{1'}\otimes\mathbf{2}$. The resulting mass matrix for the neutrinos is:
\begin{equation}
\label{w1}
m_\nu^{k_\ell=1}=\frac{2g v_u^2}{\Lambda}
\begin{pmatrix}
-Y_1 & Y_2 &\frac{g'}{2g}Y_2\\
Y_2 & Y_1 &-\frac{g'}{2g}Y_1 \\
\frac{g'}{2g}Y_2 &-\frac{g'}{2g}Y_1 &0
\end{pmatrix}\,.
\end{equation}
\item \textbf{W2} $(k_\ell=2)$: Now all the terms in (\ref{general_wein}) are allowed by the symmetry. The superpotential of the W2 reads:
\begin{multline}
\label{W2}
\mathcal{W}_\nu^{k_\ell=2}\supset \frac{g}{\Lambda}H_uH_u(D_\ell D_\ell)_\mathbf{2} Y^{(2)}_\mathbf{2}+\frac{g'}{\Lambda}H_uH_uD_\ell \ell_3 (Y^{(2)}_\mathbf{2})+\\+ \frac{g''}{\Lambda}H_uH_u(D_\ell D_\ell)_\mathbf{1} Y^{(2)}_\mathbf{1}+\frac{g_p}{\Lambda}H_uH_u\ell_3\ell_3 (Y^{(2)}_\mathbf{1})\,.
\end{multline}
This model gives rise to the following neutrino mass matrix:
\begin{multline}
\label{w2}
m_\nu^{k_\ell=2}=\frac{2g v_u^2}{\Lambda}\left[\begin{pmatrix}
-(Y_2^2-Y_1^2)&2Y_1Y_2 & \frac{g'}{2g}2Y_1Y_2 \\
2Y_1Y_2 & (Y_2^2-Y_1^2)& -\frac{g'}{2g}(Y_2^2-Y_1^2)\\ 
\frac{g'}{2g}2Y_1Y_2 & -\frac{g'}{2g}(Y_2^2-Y_1^2)& 0\end{pmatrix}
+\right.\\ \left.+\begin{pmatrix}
\frac{g''}{g}(Y_1^2+Y_2^2)&0 &0\\
0 & \frac{g''}{g}(Y_1^2+Y_2^2)&0\\
0&0& \frac{g_p}{g}(Y_1^2+Y_2^2)
\end{pmatrix}
\right]
\end{multline}
\item \textbf{W3} $(k_\ell=3)$: We have a structure similar to the W2 model:
\begin{multline}
\mathcal{W}_\nu^{k_\ell=3}\supset \frac{g}{\Lambda}H_uH_u(D_\ell D_\ell)_\mathbf{2} Y^{(3)}_\mathbf{2}+\frac{g'}{\Lambda}H_uH_uD_\ell \ell_3 (Y^{(3)}_\mathbf{2})+\\+ \frac{g''}{\Lambda}H_uH_u(D_\ell D_\ell)_\mathbf{1} Y^{(3)}_\mathbf{1}+\frac{g_p}{\Lambda}H_uH_u\ell_3\ell_3 (Y^{(3)}_\mathbf{1})\,.\end{multline}
\end{itemize}
As already mentioned, we decide not to go further than $k_\ell=3$. For example, $k_\ell=4$ involves modular multiplets of weight $8$ which come in $5$ linearly independent components, thus increasing the number of free parameters. 
\section{Numerical Simulations \label{7_num}}
Factorising two overall mass-scales\footnote{The overall mass-scales can be easily recovered by fitting $\Delta m^2_\text{atm},\Delta m^2_\text{sol}$ and the charged-lepton masses to the data, once the dimensionless observables are correctly reproduced.}($\alpha\,v_d$) and ($g\,v_u^2/\Lambda$) for the charged leptons and the neutrinos, respectively, our models will effectively be controlled by a set of dimensionless parameter ratios and the modulus $\{\tau,\beta/\alpha,\gamma/\alpha,...,g'/g,g_p/g,...\}$. We will test every model by comparing its predictions to the following set of six dimensionless observables: $\{\sin^2\theta_{12},\sin^2\theta_{13},\sin^2\theta_{23}, \\ m_e/m_\mu, m_\mu/m_\tau, r\}$ where $r\equiv \Delta m^2_\text{sol}/|\Delta m^2_\text{atm}|$. It should be emphasised that, even though we report its experimental ranges in table \ref{exp_data} for completeness, we do not include the CP phase $\delta_\text{CP}$ in 
the fit, hence the latter will be a prediction of our models. 
\begin{table}[]
\centering
\renewcommand{\arraystretch}{1.2}
\begin{tabular}{lcc} 
\toprule
Parameter$\qquad\qquad$ & \multicolumn{2}{c}{Best-fit value and $1\sigma$ range} \\ 
\midrule
$\Delta m^2_\text{sol}/(10^{-5}\text{ eV}^2)$ & \multicolumn{2}{c}{$7.36^{+0.16}_{-0.15}$} \\
& \textbf{NO} & \textbf{IO} \\
$|\Delta m^2_\text{atm}|/(10^{-3}\text{ eV}^2)$ & $2.485^{+0.023}_{-0.031}$ & $2.455^{+0.030}_{-0.025}$ \\
$r \equiv \Delta m^2_\text{sol}/|\Delta m^2_\text{atm}|$ & $0.0296\pm0.0008$ & $0.0299\pm0.0008$\\
$\sin^2\theta_{12}$ & $0.303^{+0.013}_{-0.013}$ & $0.303^{+0.013}_{-0.013}$ \\
$\sin^2\theta_{13}$ & $0.0223^{+0.0007}_{-0.0006}$ & $0.0223^{+0.0006}_{-0.0006}$ \\
$\sin^2\theta_{23}$ & $0.455^{+0.018}_{-0.015}$ & $0.569^{+0.013}_{-0.021}$ \\
$\delta_\text{CP}/\pi$ & $1.24^{+0.18}_{-0.13}$  & $1.52^{+0.14}_{-0.15}$ \\
\midrule
$m_e / m_\mu$ & \multicolumn{2}{c}{$0.0048 \pm 0.0002$} \\
$m_\mu / m_\tau$ & \multicolumn{2}{c}{$0.0565 \pm 0.0045$} \\ 
\bottomrule
\end{tabular}
\caption{\small{\it The best-fit values for the neutrino observables and their 1$\sigma$ ranges, extracted from the recent global analysis of ref.\cite{Capozzi:2021fjo}. As usual, “NO" and “IO" stand for “normal-ordering" and “inverted-ordering" of the neutrino mass eigenstates. The charged-leptons mass ratios are taken from \cite{Feruglio:2017spp}. The parameters are defined as $\Delta m^2_\text{sol} \equiv m_2^2-m_1^2$
and $\Delta m^2_\text{atm}\equiv m_3^2 - (m_1^2+m_2^2)/2$. The best-fit and 1$\sigma$ ranges for the $\delta_{CP}$ were not used in our numerical search.}
}
\label{exp_data}
\end{table}
We shall denote with $q_j(p_i)$ the observables obtained from the model with the $p_i$ set of parameters taken as input, and with $q_j^{\text{b-f}}$ the corresponding best-fit values summarised in table \ref{exp_data}. To properly explore the viable parameter regions which in modular models, as pointed out in ref.\cite{Novichkov:2018ovf}, are typically characterised by quite peculiar shapes,\footnote{In particular, approximating the parameter regions with a parallelepiped or an ellipsoid typically corresponds to an underestimation of the actual viable space.} a “figure of merit" $l(p_i)\equiv \sqrt{ \chi^2(p_i)}$ can be introduced by making use of the Gaussian approximation:\footnote{The uncertainties $\sigma_j$ are obtained by symmetrizing the 1$\sigma$ ranges given in table \ref{exp_data}.}
\begin{equation}
\label{chisq}
\chi^2(p_i)=\sum_{j=1}^6 \left(\frac{q_j(p_i)-q_j^{\text{b-f}}}{\sigma_j}\right)^2\,.
\end{equation}
Following \cite{Novichkov:2021cgl}, the exploration of the parameter regions can then be performed with a Metropolis-Hastings type of algorithm inspired from the Brownian motion of a particle in a potential that depends on $l(p_i)$. In this search, all lepton observables were calculated using the MPT \textsc{Mathematica} package\footnote{This package allows to input the (numerical) matrix entries for the charged-leptons and neutrino matrices, and returns as output $\{\theta_{12},\theta_{13},\theta_{23},\delta_\text{CP},\alpha_1,\alpha_2,\{m_1,m_2,m_3\},(m_e,m_\mu,m_\tau)\}$ i.e. the mixing angles, the CP phases (including the eventual Majorana phases $\alpha_1,\alpha_2$), the neutrino masses $m_1,m_2,m_3$ and the charged-leptons masses.} provided in ref. \cite{Antusch:2005gp}, where every parameter set $p_i$ was given as input. Since the models involve a substantial number of parameters and given their non-linearity, it is expected that one finds multiple local minima $p_i$. For every model we performed a numerical scan where a specific set $p_i$ was chosen by randomly generating all the free parameters from uniform distributions, restricting the ranges to the most promising regions. 
%%%%%%%%%%%%%%%%%%%%%%%%%%%%%%%%%%%%%%%%%%%%
\begin{table}[t]
\centering
\renewcommand{\arraystretch}{1.2}
\begin{tabular}{l|lll}
\toprule
\diagbox{\footnotesize{\text{Charged leptons}}}{\footnotesize{\text{Neutrinos}}} & Weinberg $k_\ell=2$ \\ 
\midrule 
Minimal &  (\textbf{\RomanNumeralCaps{1}}) $9~[7]$   \\ 
Hierarchical & (\textbf{\RomanNumeralCaps{2}})  $10~[8]$   \\ 
\bottomrule
\end{tabular}
\caption{\small{\it Number of free parameters in all the viable combinations of charged-leptons and neutrino models including $\text{Re}\,\tau$ and $\text{Im}\,\tau$. The label assigned for the resulting combination is displayed in the parentheses. Within square brackets, the number of effective dimensionless parameters without the two overall mass scales is highlighted. As a reminder, {\it Minimal} refers to the charged-leptons mass matrix arising from the superpotential \eqref{mcl}, {\it Hierarchical} from  \eqref{ncl}.}}
\label{allmodels}
\end{table}
\subsection{Results \label{8_res}}

Out of the three different neutrino textures illustrated in Sect.\ref{6_neutr}, \textbf{W1,W2} and \textbf{W3}, the procedure discussed above elected the Weinberg $k_\ell=2$ from \eqref{w2} as the most promising one; thus, considering the two charged-leptons models (Minimal (\ref{mcl}) and Hierarchical (\ref{ncl})), we ended up with two distinct and well performing models for lepton masses and mixing, that  we call model \textbf{\RomanNumeralCaps{1}} and model \textbf{\RomanNumeralCaps{2}} respectively, which are summarised in table \ref{allmodels}. 
Overall, model \textbf{\RomanNumeralCaps{1}} is described by seven free parameters and two mass scales ($\text{Re}\, \tau,~\text{Im}\,\tau,~\beta/\alpha,~\gamma/\alpha,~g'/g,~g''/g,~g_p/g$ and $v_d\,\alpha$,~$v_u^2\, g / \Lambda$) while model \textbf{\RomanNumeralCaps{2}} has eight  free parameters and two mass scales 
($\text{Re}\, \tau,~\text{Im}\,\tau,~\beta/\alpha, \\~\gamma/\alpha,~\alpha_D/\alpha,~g'/g,~g''/g,~g_p/g$ and $v_d\,\alpha$,~$v_u^2\, g / \Lambda$).
The results of our numerical scan are listed in table \ref{MCW2_table} (model \textbf{\RomanNumeralCaps{1}}), and in table \ref{NCW2_table} (model \textbf{\RomanNumeralCaps{2}}). As can be seen, the agreement with data is excellent. It must be noted that, given the CP-invariance imposed on the models, under the transformation\footnote{Given the fact that all the couplings are real, and that the q-expansions (\ref{qexp}) have real coefficients, the transformation $\tau\to-\tau^*$ is equivalent to a complex conjugation of the mass matrices. The only observables that are affected in the mass matrices are the CP phases.} $\tau\to-\tau^*$ we obtain the same observables while the CP phases ($\delta_\text{CP},\alpha_1,\alpha_2$) flip their signs. Since the only complex parameter in our models is $\text{Re}\,\tau$, every viable set of parameters comes in pairs distinguished only by $\pm\text{Re}\,\tau$, which in the fundamental domain $\mathcal{D}$ (figure \ref{ffund}) appears as a reflection with respect to the imaginary axis. \newline

%%%%%%%%%%%%%%%%%%%%%%%%%
%%%%%%%%%%%%%%%%%%%%%%%%%%%%%%%%%%%%%%%%%%%%%%%
%%%%%%%%%%%%%%%%%%%%%%%%%%%%%%%%%%%%%%%%%%%%%%%
%%%%%%%%%%%%%%%%%%%%%%%%%%%%%%%%%%%%%%%%%%%%%%%
%%%%%%%%%%%%%%%%%%%%%%%%%%%%%%%%%%%%%%%%%%%%%%%

%%%%%% TABLES TABLES TABLES %%%%%%%%%%%

%%%%%%%%%%%%%%%%%%%%%%%%%%%%%%
\begin{table}
\centering
\renewcommand{\arraystretch}{1}
\begin{tabular}{c|ccc}
   \toprule
  & Best-fit and $1\sigma$ range \\
  \midrule
  $\text{Re}\, \tau$ & $\pm 0.0895_{-0.0055}^{+0.0032}$  \\
  $\text{Im}\,\tau$ & $1.697_{-0.016}^{+0.023}$  \\
  $\beta/\alpha$ & $14.33_{-0.38}^{+0.58}$  \\
  $\gamma/\alpha$ & $17.39_{-0.87}^{+1.38}$  \\
  $g'/g$ & $31.57_{-10.29}^{+27.59}$  \\
  $g''/g$ & $7.17_{-2.36}^{+6.36}$ \\
  $g_p/g$ & $8.51_{-3.03}^{+7.99}$ \\
  $v_d\,\alpha$ [MeV] & 102.14 \\
  $v_u^2\, g / \Lambda$ [eV] & 0.47 \\
  \midrule
  $\sin^2 \theta_{12}$ & $0.300_{-0.006}^{+0.013}$  \\
  $\sin^2 \theta_{13}$ & $0.0223_{-0.0006}^{+0.0004}$   \\
  $\sin^2 \theta_{23}$ & $0.452_{-0.009}^{+0.015}$ \\
  $r$ & $0.0295_{-0.0006}^{+0.0007}$ \\
  $m_e/m_{\mu}$ & $0.0048_{-0.0002}^{+0.0001}$ \\
  $m_{\mu} / m_{\tau}$ & $0.0578_{-0.0040}^{+0.0023}$   \\
  \midrule
  Ordering & \textbf{NO} \\
  $\delta/\pi$ & $\pm 1.594_{-0.010}^{+0.007}$   \\
  $m_1$ [eV] & $0.0182_{-0.0014}^{+0.0018}$ \\
  $m_2$ [eV] & $0.0201_{-0.0013}^{+0.0017}$ \\
  $m_3$ [eV] & $0.0537_{-0.0005}^{+0.0006}$  \\
  $\textstyle \sum_i m_i$ [eV] & $0.092_{-0.003}^{+0.004}$  \\
  $|m_{\beta\beta}|$ [meV] & $18.89_{-1.47}^{+1.90}$   \\
   $m_{\beta}^{\text{eff}}$ [meV] & $20.26_{-1.30}^{+1.69}$  \\
  $\alpha_{1}/\pi$ & $\pm 1.124_{-0.017}^{+0.014}$  \\
  $\alpha_{2}/\pi$ & $\pm 0.949_{-0.005}^{+0.005}$  \\
    \midrule
$d_\text{FT}$ & $12.2$ \\
  \midrule
  $\chi^2_\text{min}$ & 0.16 \\
  \bottomrule
\end{tabular}
\caption{
\small{\it Best fit values and $1\sigma$ range for parameters and observables in model \textbf{\RomanNumeralCaps{1}}  ($7$ free parameters and $2$ mass scales). $\alpha_1,\alpha_2$ are the Majorana phases; $|m_{\beta\beta}|$ is the effective Majorana mass for the neutrinoless double-beta decay; $m_{\beta}^{\text{eff}}$ is the effective neutrino mass in the single beta decay, probed by tritium decay experiments.}}
\label{MCW2_table}
\end{table}
%%%%%%%%%%%%%%%%%%%%%%%%%%%%%%

%%%%%%%%%%%%%%%%%%%%%%%%%%%%%%
\begin{table}
\centering
\renewcommand{\arraystretch}{1}
\begin{tabular}{c|ccc}
   \toprule
  & Best-fit and $1\sigma$ range \\
  \midrule
  $\text{Re}\, \tau$ & $\pm 0.090_{-0.004}^{+0.004}$  \\
  $\text{Im}\,\tau$ & $1.688_{-0.018}^{+0.015}$  \\
  $\beta/\alpha$ & $1.03_{-0.04}^{+0.04}$  \\
  $\gamma/\alpha$ & $1.26_{-0.08}^{+0.12}$  \\
  $\alpha_D/\alpha$ & $1.33_{-1.05}^{+1.51}$  \\
  $g'/g$ & $41.9_{-12.8}^{+73.7}$  \\
  $g''/g$ & $9.55_{-2.91}^{+16.81}$ \\
  $g_p/g$ & $11.5_{-3.8}^{+21.2}$ \\
  $v_d\,\alpha$ [MeV] & 1404.6 \\
  $v_u^2\, g / \Lambda$ [eV] & 0.35 \\
  \midrule
  $\sin^2 \theta_{12}$ & $0.305_{-0.015}^{+0.009}$  \\
  $\sin^2 \theta_{13}$ & $0.0222_{-0.0006}^{+0.0007}$   \\
  $\sin^2 \theta_{23}$ & $0.454_{-0.008}^{+0.007}$ \\
  $r$ & $0.0295_{-0.0007}^{+0.0007}$ \\
  $m_e/m_{\mu}$ & $0.0048_{-0.0002}^{+0.0002}$ \\
  $m_{\mu} / m_{\tau}$ & $0.0570_{-0.0048}^{+0.0034}$   \\
  \midrule
  Ordering & \textbf{NO} \\
  $\delta/\pi$ & $\pm 1.597_{-0.006}^{+0.009}$   \\
  $m_1$ [eV] & $0.0174_{-0.0014}^{+0.0011}$ \\
  $m_2$ [eV] & $0.0194_{-0.0012}^{+0.0010}$ \\
  $m_3$ [eV] & $0.0535_{-0.0004}^{+0.0004}$  \\
  $\textstyle \sum_i m_i$ [eV] & $0.090_{-0.003}^{+0.002}$  \\
  $|m_{\beta\beta}|$ [meV] & $18.14_{-1.48}^{+1.17}$   \\
   $m_{\beta}^{\text{eff}}$ [meV] & $19.60_{-1.25}^{+1.02}$  \\
  $\alpha_{1}/\pi$ & $\pm 1.129_{-0.013}^{+0.019}$  \\
  $\alpha_{2}/\pi$ & $\pm 0.946_{-0.004}^{+0.004}$  \\
    \midrule
$d_\text{FT}$ & $11.2$ \\
  \midrule
  $\chi^2_\text{min}$ & 0.074 \\
  \bottomrule
\end{tabular}
\caption{
\small{\it Same as table \ref{MCW2_table} but for model \textbf{\RomanNumeralCaps{2}} ($8$ free parameters and $2$ mass scales).}}
\label{NCW2_table}
\end{table}
%%%%%%%%%%%%%%%%%%%%%%%%%%%%%%%%%%%%%%%%%%%%%%%%%%%%%%%%%%%%%%%%%%%%%%%%%%%%%%%%%%%%%%%%%%%%%%%%%%%%%%%%%%%%%%%%%%%%%%%%%%%%%%

%%%%%%%%%%%%% PLOT%%%%%%%%%%%%% PLOT %%%%%%%% PLOT 
%%%%%%%%%%%%%%%%%%%%%%%%%%%%%%%%%%%%%%%%%%%%%%%
%%%%%%%%%%%%%%%%%%%%%%%%%%%%%%%%%%%%%%%%%%%%%%%
%%%%%%%%%%%%%%%%%%%%%%%%%%%%%%%%%%%%%%%%%%%%%%%
%%%%%%%%%%%%%%%%%%%%%%%%%%%%%%%%%%%%%%%%%%%%%%%
%%%%%%%%%%%%%%%%%%%%%%%%%%%%%%%%%%%%%%%%%%%%%%%
\begin{figure}
  \begin{center}
    \includegraphics[scale=0.25]{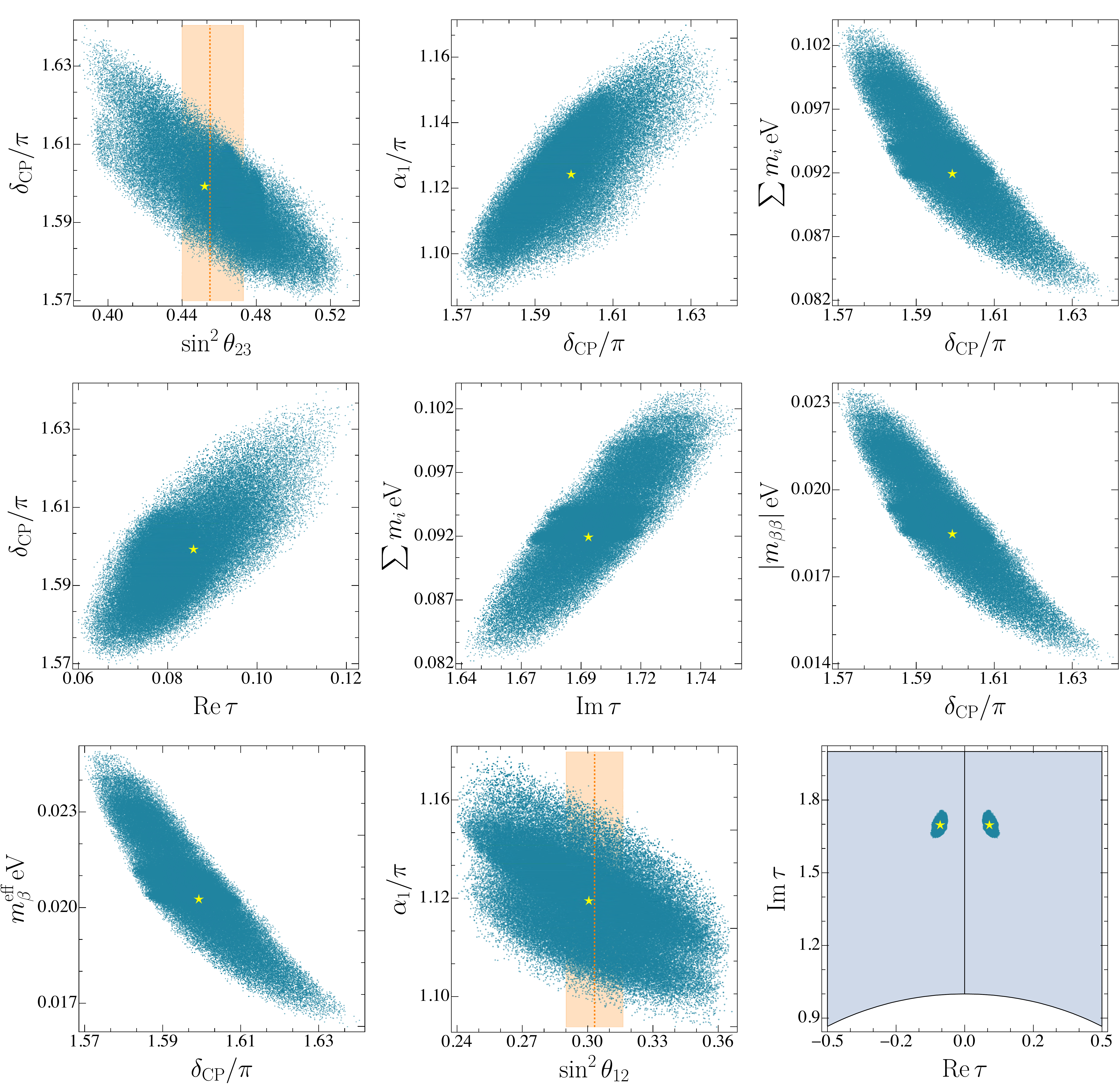}
\caption{\small{\it Correlations between pair of observables and parameters in model \textbf{\RomanNumeralCaps{1}}. The plotted points correspond to all the points accepted by the exploration algorithm described in ref. \cite{Novichkov:2021cgl}. The points coming from the parameter-set with the minimum $\chi^2$ are highlighted in yellow. The orange lines and bands are the best-fit and $1\sigma$ range values from ref.\cite{Capozzi:2021fjo}.}}
   \label{IIcorrelations}
  \end{center}
  \end{figure}

\begin{figure}
  \begin{center}
    \includegraphics[scale=0.25]{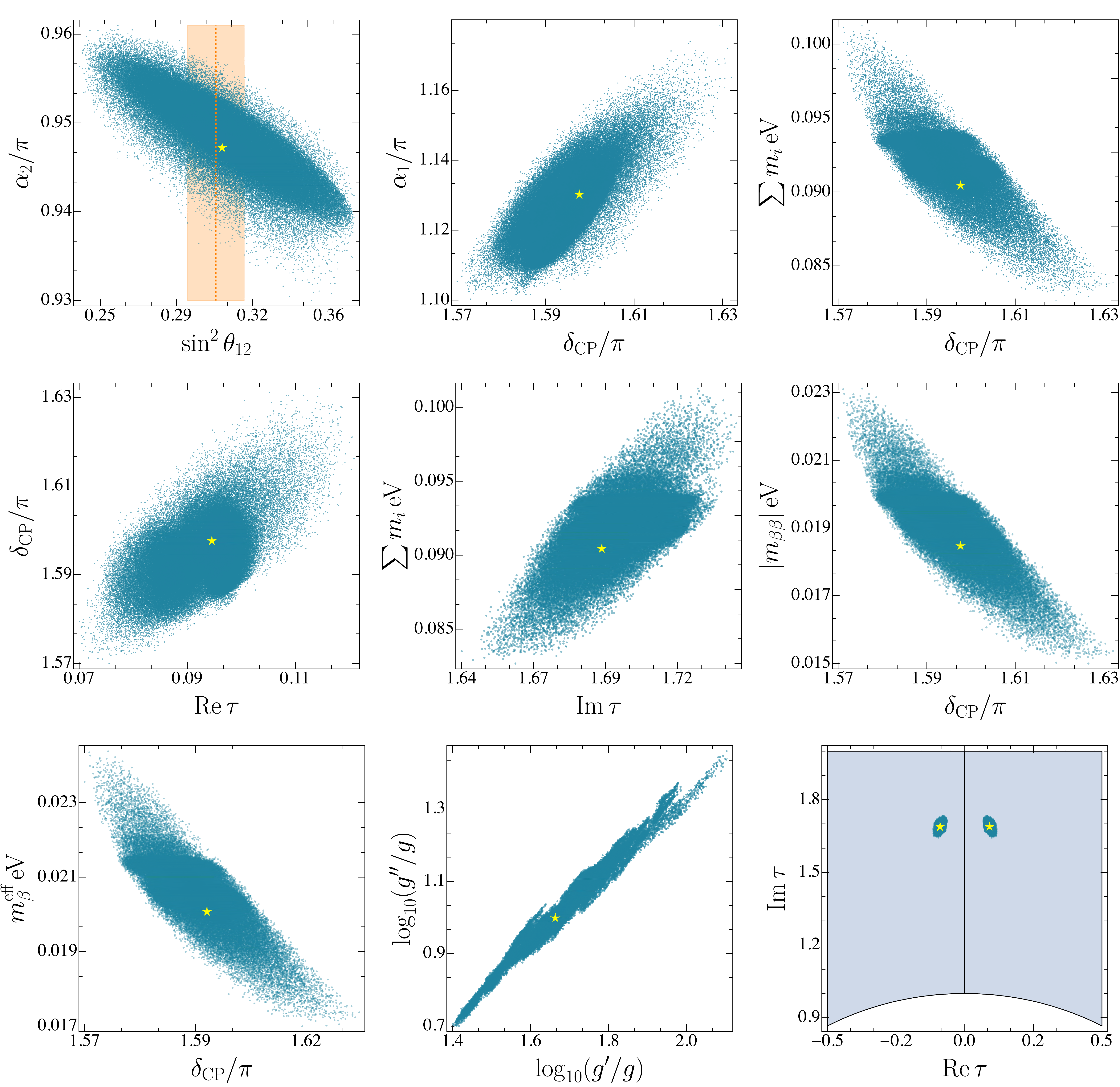}
\caption{\small{\it Same as figure \ref{IIcorrelations} but for model \textbf{\RomanNumeralCaps{2}}. }}
   \label{IVcorrelations}
  \end{center}
  \end{figure}

%%%%%%%%%%%%%%%%%%%%%%%%%%%%%%%%%%%%%%%%%%%%%%%%%%%%%%%%%%%%%%%%%%%%%%%%%%%%%%%%%%%%%%%%%%%%%%%%%%%%%%%%%%%%%%%%%%%%%%%%%%%%%%
%%%%%%%%%%%%%%%%%%%%%%%%%%%%%%%%%%%%%%%%%%%%%%%%%%%%%%%%%%%%%%%%%%%%%%%%%%%%%%%%%%%%%%%%%%%%%%%%%%%%%%%%%%%%%%%%%%%%%%%%%%%%%%

% \vspace{5 mm}
\vspace{1mm}

Following \cite{Criado:2018thu}, the $1\sigma$ ranges reported in the tables were obtained by the calculation of the one-dimensional projections of the $\chi^2$, and correspond to the increase by one unit from $\chi^2_\text{min}$. Moreover, as a quantitative measure of the amount of fine-tuning in a given model, we included in each table the normalised Altarelli-Blankenburg measure $d_\text{FT}$ from ref. \cite{Altarelli:2010at}:
\begin{equation}
\label{alta_blank}
d_\text{FT}=\frac{\sum_i\left|\frac{\text{par}_i}{\delta\text{par}_i}\right|}{\sum_i\left|\frac{\text{obs}_i}{\sigma_i}\right|}\,.
\end{equation}
In the numerator there appears the sum of the absolute ratios between the best-fit values of each parameter and $\delta\text{par}_i$, where the latter is defined as the shift of the parameter from its best-fit value that increases the $\chi^2$ by one unit while keeping all the other parameters fixed.\footnote{The intuition behind it is that we consider “fine-tuned" a model in which it takes a minimal variation of the free parameters to produce a large variation in the $\chi^2$.} This quantity is normalised by the sum of the absolute ratios between the observables and their uncertainties, taken from the input data of table \ref{exp_data}. As it can be seen in tables \ref{MCW2_table} and \ref{NCW2_table}, the fine-tuning in both models returns very similar values, as can be expected from the fact that the two models are distinguished by only one free parameter. In addition, since its value is of ${\cal{O}}$(10),
we can safely argue that the models do not have an intrinsic fine-tuning much larger than the one already present in the data.
\newline
 
% \vspace{2mm}
A set of interesting correlations among model parameters and observables has been reported 
in figure \ref{IIcorrelations} for model \textbf{\RomanNumeralCaps{1}} and figure \ref{IVcorrelations} for model \textbf{\RomanNumeralCaps{2}}.
As a first observation, we can see that the free parameter $g'/g$ (that we explicitly show in figure \ref{IVcorrelations} only, having a very similar behavior for \textbf{\RomanNumeralCaps{1}} also) can vary in a large range and still provide good agreement with data in both models. This translates in a large 1$\sigma$ range, as evident from tables \ref{MCW2_table} and \ref{NCW2_table}. Notice that this parameter enters in the Weinberg superpotential, and a similar behaviour was pointed out by Novichkov et al. in \cite{Novichkov:2020eep} in the context of the $S_4'$ group.\newline
During the execution of our numerical scan, we also noted that viable values for $g'$ and $g''$ are linearly correlated with a slope near $1/4$. This feature can be understood analytically. First, consider the neutrino mass matrix in eq \eqref{w2} with ${g''}\sim {g_p}$:
\begin{multline}
\label{minimal_w2}
m_\nu^{k_\ell=2}=\frac{2g v_u^2}{\Lambda}\left[\begin{pmatrix}
-(Y_2^2-Y_1^2)&2Y_1Y_2 & \frac{g'}{2g}2Y_1Y_2 \\
2Y_1Y_2 & (Y_2^2-Y_1^2)& -\frac{g'}{2g}(Y_2^2-Y_1^2)\\ 
\frac{g'}{2g}2Y_1Y_2 & -\frac{g'}{2g}(Y_2^2-Y_1^2)& 0\end{pmatrix}
+\right.\\ \left.+\frac{g''}{g}\begin{pmatrix}
(Y_1^2+Y_2^2)&0 &0\\
0 &(Y_1^2+Y_2^2)&0\\
0&0&(Y_1^2+Y_2^2)
\end{pmatrix}
\right]\,.
\end{multline}
Factorising $Y_1^2g'/g$, we explore the limit $g'/g \gg 1$, where this matrix can be approximated in the following way (apart from overall factors):
\begin{equation}
\label{approx_mw2}
m_\nu^{k_\ell=2}\sim \begin{pmatrix}
 G(1+\epsilon^2)&0 &\epsilon\\
 0& G(1+\epsilon^2)& (1-\epsilon^2)/2\\
\epsilon &(1-\epsilon^2)/2 &G(1+\epsilon^2)
\end{pmatrix}\,,
\end{equation}
where we called $G\equiv g''/g'$ and $\epsilon\equiv Y_2/Y_1$ is the same expansion parameter used in section \ref{hierac_mod}. Diagonalising $m_\nu^\dagger m_\nu$ we find the approximated neutrino masses (assuming $G>0$) 
\begin{align}
\label{neumasses_limit}
\centering
&m_1^2\sim G^2(1+2|\epsilon|^2)+\mathcal{O}(\epsilon^3)\\
&m_2^2\sim (1/2-G)^2(1+2|\epsilon|^2)+\mathcal{O}(\epsilon^3)\\
&m_3^2\sim (1/2+G)^2(1+2|\epsilon|^2)+\mathcal{O}(\epsilon^3)\,,
\end{align}
and it is easy to show that the ratio $r\equiv \Delta m^2_\text{sol}/\Delta m^2_\text{atm}$ can be expressed as:
\begin{equation}
\label{ratio_limit}
r=\frac{1/4-G}{2G}+\mathcal{O}(\epsilon^3)\,.
\end{equation}
From this expression we find that $r\approx 1/30$ for $G\approx 0.234$, as observed in the numerical scans. \newline
\vspace{1mm}

As for the physical observables, a close inspection of tables \ref{MCW2_table} and \ref{NCW2_table} and figures \ref{IIcorrelations} and \ref{IVcorrelations} reveal 
interesting model predictions. In particular, both models predict a normal ordered neutrino spectrum. This is in agreement with the conclusions from the global analysis on recent oscillation data from ref.\cite{Capozzi:2021fjo}, where the normal ordering is actually favored at a global $\sim 3\sigma$ level. In addition, both models predict:

\begin{itemize}
\item  a CP-violating phase $\delta_\text{CP}\sim \pm1.6\pi$, which in general lies in a pretty narrow interval $\delta_\text{CP}/\pi\sim\pm[1.57,1.63]$ as can be seen from the plots in figures \ref{IIcorrelations} and \ref{IVcorrelations}. It is remarkable that the predicted value of $\sim 1.6\pi$ is in agreement (within the $2\sigma$ range) with the global analysis of oscillation data \cite{Capozzi:2021fjo} for the NO;
\item values of the sum of neutrino masses $\sum_i m_i$ around $0.090\ev$, which is compatible with the present upper bound of $0.115\,\text{eV}$ ($95\,\%\,\text{C.L.}$), see \cite{eBOSS:2020yzd};
\item the Majorana effective mass for the neutrinoless double-beta decay to lie around $\sim\,20\,\text{meV}$, not too far from the recent KamLAND-Zen upper bound $|m_{\beta\beta}|<(36-156)\,\text{meV}$ \cite{KamLAND-Zen:2022tow}; 
\item both Majorana phases $\alpha_1,\alpha_2$  in narrow regions around $\pm 1.13\pi, \pm 0.95\pi$, respectively. 
\end{itemize}
Our predictions for $\delta_\text{CP}$ can be tested in forthcoming oscillation experiments (DUNE \cite{Strait2016}, and T2HK \cite{Hyper-KamiokandeProto-:2015xww}), whereas the Majorana effective mass $|m_{\beta\beta}|$ for the neutrinoless double-beta decay can be probed by future ton-scale experiments (e.g. nEXO \cite{nEXO:2018ylp}); both classes of experiments will help to validate or discard our models.

\section{Conclusions}
\label{concl}
In this work we found two viable models for lepton masses and mixing based on the modular group $\Gamma_2\cong S_3$ and gCP symmetry, providing a clear improvement with respect to the only existing modular $S_3$ (SUSY) model from ref. \cite{Kobayashi:2018vbk}. For the first time, $S_3$ lepton constructions (models \textbf{\RomanNumeralCaps{1}} and \textbf{\RomanNumeralCaps{2}}) have been realised without the aid of beyond Standard Model fields besides the modulus, and with the fewest number of free parameters (nine in the minimalist model \textbf{\RomanNumeralCaps{1}}). For the VEV of $\tau$ inside the fundamental domain \eqref{fun_domain}, the charged-leptons mass hierarchy is accounted for by symmetry arguments with the careful assignments of irreps and modular charges of the superfields. In model \textbf{\RomanNumeralCaps{2}} the charged-leptons mass pattern is $\sim m_\tau(1,|Y_1|,|Y_1|^3)$ with the small parameter $|Y_1|\approx 7/100$ substantially arising from the ad-hoc normalisation of the modular forms \eqref{multiplets1} with the choice \eqref{cinorm}. As a result, the Higgs couplings of charged-leptons come with $\mathcal{O}(1)$ ratios. In model \textbf{\RomanNumeralCaps{1}} the mass pattern is $\sim m_\tau(1,|Y_1|,|Y_1|^2)$, thus in this case the corresponding couplings ratios have to be of order $\mathcal{O}(10)$. In both models, neutrino masses are then generated through Weinberg operators following the assignments made for the charged-leptons sector. 
We performed a numerical scan and found an excellent fit to neutrino mixing data, with the prediction of a normal ordered spectrum, narrow ranges for the Dirac CP violation phase $\delta_\text{CP}/\pi\in [1.57,1.63]$ and for the Majorana phases $\alpha_1/\pi\in [1.10,1.17]\,,\,\alpha_2/\pi\in [0.93,0.96]$. All the predictions for the sum of neutrino masses, neutrinoless double-beta decay and tritium decay effective masses are compatible with present experimental bounds. Some of these predictions can be tested by forthcoming neutrino experiments, in particular the ordering, the CP-violating phase and the effective mass for the neutrinoless double-beta decay. \newline
Our results reopen the interest for a unification of quark-lepton flavour theories under the $\Gamma_2\cong S_3$ group, in schemes with a smaller number of free parameters compared to the Standard Model.

Different sources of corrections should be considered. Since in modular models the masses and mixing angles are given at a high-energy scale, they are sensible to corrections due to the renormalization group running, and since these models are also formulated in a supersymmetric context they may be sensible to SUSY-breaking effects. For the latter, it was shown in ref. \cite{Criado:2018thu} that the subsequent corrections can be made negligible provided some mild conditions on the scale $M$ at which the SUSY-breaking sector communicates with the visible sector. For the renormalization group running effects, in ref. \cite{Criado:2018thu} these were studied for a model predicting a normal ordering and with a mass of the lightest neutrino $m_\text{lightest}\sim 0.01\,\text{eV}$, and it was shown that they are negligible. As pointed out by Antusch et al. in ref. \cite{Antusch:2003kp}, the running of neutrino masses and mixing depend mostly on the ratio of the VEVs of the two Higgs fields ($\braket{H_u}=v_u$, $\braket{H_d}=v_d$) defined as $\tan\beta\equiv v_u/v_d$ and on $m_\text{lightest}$. For normal ordering the running effects generally decrease for lower $\tan\beta$ and lower\footnote{For the neutrino masses the running is negligible if we do not have a quasi-degenerate spectrum, which is our case since $m_\text{lightest}<\sqrt{|\Delta m^2_\text{atm}|}\sim 0.05\,\text{eV}$ \cite{Ereditato:2018eqf}. In both models we obtained a normal ordering and $m_\text{lightest}\sim 0.02\,\text{eV}$.} $m_\text{lightest}$. Since our phenomenology is similar to what was obtained in ref. \cite{Criado:2018thu}, we conclude that renormalization group corrections are negligible in our case. As a final source of corrections, one could argue that non-minimalistic contributions to the Kähler potential \eqref{kahler} are not forbidden by modular symmetry \cite{Chen:2019ewa}. In the absence of a definitive UV construction that excludes such contributions, there is no a priori reason to do so in a bottom-up approach, hence the problem is an open one \cite{Feruglio:2019ybq}. In that regard, it was shown that in a “top-down" construction the deviations from the minimalistic form \eqref{kahler} can be made smaller than the experimental errors of neutrino parameters \cite{Chen:2021prl}.
\newline
\vspace{1mm}

\section*{Acknowledgements}
We thank J.Penedo and A.Titov for very useful discussions.

%%%%%%%%%%%%%%% APPENDICE APPENDICE APPENDICE %%%%%%%%%%%%%%%%%%%%%%%%
%%%%%%%%%%%%%%% %%%%%%%%%%%%%%% %%%%%%%%%%%%%%% %%%%%%%%%%%%%%% 
\begin{appendices}
\section{$S_3$ and its representations\label{appendix}}
We summarise here the main results from the theory of non-Abelian discrete groups (outlined in ref. \cite{Ishimori:2012zz}) applied to $S_3$. This is the group of all possible permutations of three objects $(x_1,x_2,x_3)$. These permutations are given by $3!=6$ group elements defined as:
\begin{equation}
\label{elems}
\begin{split}
e\,:\,(x_1,x_2,x_3)\to(x_1,x_2,x_3)\\
a_1\,:\,(x_1,x_2,x_3)\to(x_2,x_1,x_3)\\
a_2\,:\,(x_1,x_2,x_3)\to(x_1,x_3,x_2)\\
a_3\,:\,(x_1,x_2,x_3)\to(x_3,x_2,x_1)\\
a_4\,:\,(x_1,x_2,x_3)\to(x_3,x_1,x_2)\\
a_5\,:\,(x_1,x_2,x_3)\to(x_2,x_3,x_1)
\end{split}
\end{equation}
It can be shown that, for example, the two simple transpositions $a_1$ and $a_3$ can generate all the other permutations. Calling $a\equiv a_3$, $b\equiv a_1$, the group is given by
\begin{equation}
\label{s3elems}
S_3=\{e,a,b,ab,ba,bab\}\,,
\end{equation}
and its pictorial illustration as the symmetry of an equilateral triangle is shown in figure \ref{s3labb}. 
\renewcommand{\thefigure}{\Alph{section}.\arabic{figure}}
\setcounter{figure}{0}
\begin{figure}[t]
  \begin{center}
    \includegraphics[scale=0.2]{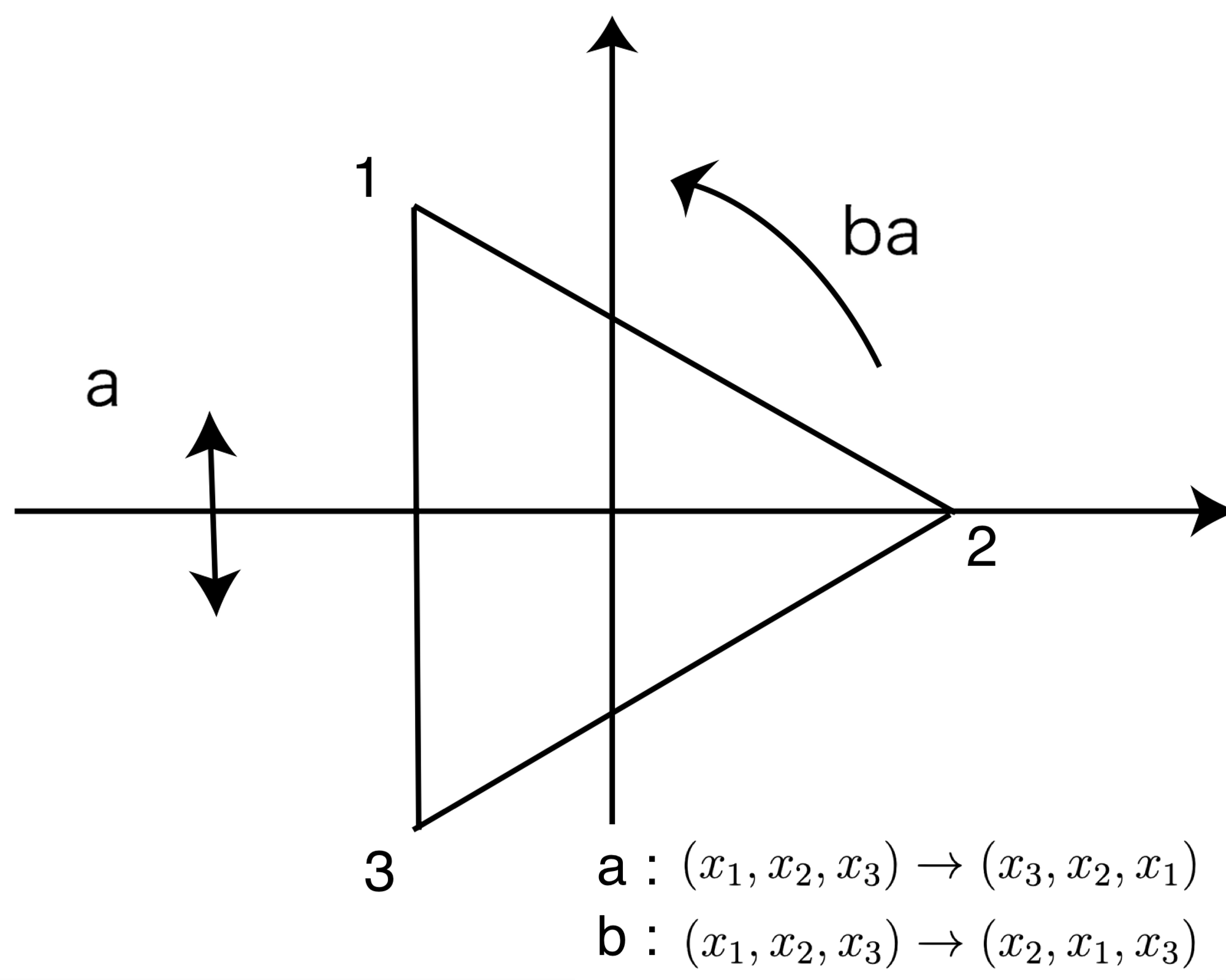}
    \caption{\small{\it The generators of the $S_3$ group: two reflections around the bisectors of an equilateral triangle. Their combined action $ba$ is a counter-clockwise rotation of $2\pi/3$.}}
    \label{s3labb}
  \end{center}
  \end{figure}
The $S_3$ group can be partitioned into three conjugacy classes\footnote{Given a group $\mathcal{G}$, the conjugacy class of an element $a\in\mathcal{G}$ is given by the set:
$$
C(a)=\{gag^{-1}\,,\,\forall g\in\mathcal{G}\}
$$} labelled by $C_n$ (here $n$ is the number of group elements inside each class)
\begin{equation}
\label{partts3}
C_1=\{e\}\quad,\quad C_2=\{ab,ba\}\quad,\quad C_3=\{a,b,bab\}\,,
\end{equation}
these classes have orders\footnote{The “order" of an element $a\in\mathcal{G}$ is the lowest integer $h$ such that $a^h=e$ where $e$ is the identity. The order is constant inside a conjugacy class: $$(gag^{-1})^h=(gag^{-1})(gag^{-1})...(gag^{-1})=ga^hg^{-1}=geg^{-1}=gg^{-1}=e$$}  $h[C_1]=1$, $h[C_2]=3$, $h[C_3]=2$. Furthermore, it can be shown \cite{Ishimori:2012zz} that the number of irreducible representations of a discrete group must be equal to the number of its conjugacy classes. Hence, we expect three irreducible representations for $S_3$. To find them, we will need some important theorems first. More information on the irreducible representations of $S_3$ is acquired through character theory.\footnote{Given a $d_\alpha$-dimensional representation $D_\alpha(g)$ for $g\in\mathcal{G}$, the character $\chi_D(g)$ is defined as the trace of the representation matrix: 
$$\chi_D(g)=\text{Tr}(D(g))=\sum_{i=1}^{d_\alpha}D(g)_{ii}$$
From the cyclic property of the trace, we have that $\chi_D(gag^{-1})=\chi_D(g^{-1}ga)=\chi_D(a)$, that is, the character is constant inside a conjugacy class.} Given two representations $D_\alpha(g)$ and $D_\beta(g)$, their characters satisfy the orthogonality relation:
\begin{equation}
\label{ortcar}
\sum_{g\in\mathcal{G}}\chi_{D_\alpha}(g)^*\chi_{D_\beta}(g)=N\delta_{\alpha\beta}\,,
\end{equation}
where $N$ is the number of group elements. Another orthogonality relation can be obtained by summing over all the irreducible representations of the group:
\begin{equation}
\label{ortcarrr}
\sum_{\alpha}\chi_{D_\alpha}(C_i)^*\chi_{D_\alpha}(C_j)=\frac{N\delta_{ij}}{n_i}\,,
\end{equation}
where $C_i$ are the conjugacy classes and $n_i$ is the number of elements inside each class. Let us suppose that we have $m_n$ irreducible representations of dimension $n$: in all of these irreps the identity is always represented by a $n\times n$ identity-matrix, hence for $C_1\equiv \{e\}$ and $\chi_{D_\alpha}(C_1)=n$ (\ref{ortcarrr}) yields
\begin{equation}
\label{sssium}
\sum_{\alpha}|\chi_{D_\alpha}(C_1)|^2=\sum_n m_n n^2=m_1+4m_2+9m_3+16m_4+...=N\,,
\end{equation}
with the dimension $n$ increasing. But the total number of irreducible representations (which is $m_n$ summed over the index $n$) must be equal to the number of conjugacy classes. In the case of $S_3$ the two conditions translate to:
\begin{align}
\label{conds3}
m_1+4m_2+9m_3+...=6\\
m_1+m_2+m_3+...=3\,.
\end{align}
Since $m_n$ is a positive integer, these two are satisfied only by $(m_1=2, m_2=1)$: this means that we have two singlet representations and one doublet. We denote them by $\mathbf{1},\mathbf{1'},\mathbf{2}$. Exploiting (\ref{ortcar}) and (\ref{ortcarrr}) we obtain the characters for the conjugacy classes (and, as a consequence, the character for every element of $S_3$). The results are summarised in table \ref{tabss3}.
\begin{table}[]
\centering
\def\arraystretch{1.7}
\begin{tabular}{|c|c|c|c|c|}
\hline
 & $h$&$\chi_\mathbf{1}$ & $\chi_\mathbf{1'}$ & \multicolumn{1}{l|}{$\chi_\mathbf{2}$} \\ \hline
$C_1$ & $1$          & $1$    &  $1$        &   2           \\ \hline
$C_2$ & $3$            & $1$      &    $1$      &    -1       \\ \hline
$C_3$ & $2$           & $1$      &    $-1$     &     0       \\ \hline
\end{tabular}
\caption{\small{{\it Character table for the group $S_3$ in the irreducible representations $\mathbf{1},\mathbf{1'},\mathbf{2}$ for the three conjugacy classes of order $h$.}}}
\label{tabss3}
\end{table}
The singlet $\mathbf{1}$ and the pseudo-singlet $\mathbf{1'}$ representations are simply given by the characters $\chi_\mathbf{1},\chi_\mathbf{1'}$. We see that $\mathbf{1'}$ is called pseudo-singlet since it differs from $\mathbf{1}$ only for the representation of the elements inside $C_3$. If $y_1$ transforms as a pseudo-singlet of $S_3$, the action of the group generators $a$ and $b$ is:
$$a\,:\, y_1\quad\to\quad -y_1$$
$$b\,:\, y_1\quad\to\quad -y_1\,.$$
We now construct the basis for the doublet representation. For the generators of $S_3$ we have that $a,b\in C_3$, hence $a^2=b^2=e$. From table \ref{tabss3} we also learn that their $2\times2$ matrix representation must be traceless. If we want to use a real basis, then the most obvious choice is:
\begin{equation}
\label{geners}
a=\begin{pmatrix}
1&0\\
0&-1
\end{pmatrix}\quad,\quad b=\begin{pmatrix}
\sqrt{1-m_2m_3}&m_2\\
m_3&-\sqrt{1-m_2m_3}
\end{pmatrix}\,,
\end{equation}
for $m_2,m_3\in\mathbb{R}$ such that $m_2m_3\neq0$ and $m_2m_3\le 1$. A common parametrization is given by $m_2=m_3=\sin\theta$ with $\theta\in[0,2\pi]$. With this choice we have, for example, that the element $ba$ is given by:
$$
ba=\begin{pmatrix}
\cos\theta&\sin\theta\\\sin\theta&-\cos\theta
\end{pmatrix}\begin{pmatrix}
1&0\\0&-1
\end{pmatrix}=\begin{pmatrix}
\cos\theta&-\sin\theta\\\sin\theta&\cos\theta
\end{pmatrix}\,.
$$
Since $ba\in C_2$ we read from table \ref{tabss3} that $\text{Tr}(ba)=-1$, thus:
$$2\cos\theta=-1\,,$$
which is satisfied by $\theta=2\pi/3,4\pi/3$. To follow the convention from \cite{Ishimori:2012zz}, we choose $\theta=4\pi/3$. The generators\footnote{We call them $\rho(S)$ and $\rho(T)$ to match their $\Gamma_2$ counterparts.} (\ref{geners}) are then given by:
\begin{equation}
\label{genersss}
\begin{split}
\rho(S)=\frac{1}{2}\begin{pmatrix}
-1&-\sqrt{3}\\
-\sqrt{3}&1
\end{pmatrix}
\quad,\quad
\rho(T)=\begin{pmatrix}
1&0\\
0&-1
\end{pmatrix}\\ \\
(\rho(S))^2=(\rho(T))^2=(\rho(S)\rho(T))^3=\mathbbm{1}\quad\,.
\end{split}
\end{equation}
We see that in the chosen (real) basis, the two generators are symmetric matrices. \newline
\vspace{1mm}
\section{Clebsch-Gordan coefficients \label{cgc}}
We now focus on the tensor decomposition rules of $S_3$. If we indicate with $y_i$ some pseudo-singlets for $i=1,2,3...$, and with $\psi_{1,2},\varphi_{1,2}$ the components of two doublets, the tensor decomposition rules are given by:
\begin{align}
\begin{split}
\mathbf{1'}\otimes\mathbf{1'}=\mathbf{1}\quad \sim\quad  y_1y_2\,,\\
 \mathbf{1'}\otimes\mathbf{2}=\mathbf{2}\quad \sim\quad \begin{pmatrix}-y_1\psi_2 \\ y_1\psi_1\end{pmatrix}
\end{split}\,,
\end{align}

\begin{equation}
\label{cbcofff}
\mathbf{2}\otimes\mathbf{2}=\mathbf{1}\oplus\mathbf{1'}\oplus\mathbf{2}\,\begin{cases}
\mathbf{1}\quad\sim\quad \psi_1\varphi_1+\psi_2\varphi_2\\ \\
\mathbf{1'}\quad\sim\quad \psi_1\varphi_2-\psi_2\varphi_1\\ \\
\mathbf{2}\quad\sim\quad\begin{pmatrix} \psi_2\varphi_2-\psi_1\varphi_1\\ 
\psi_1\varphi_2+\psi_2\varphi_1
\end{pmatrix}\,.
\end{cases}
\end{equation}
We see that the Clebsch-Gordan coefficients are real. To the knowledge of the present authors, in the absence of a definitive top-down approach it is still unclear how these decompositions should be normalised. In this paper we arbitrarily set to $1$ all the coefficients in front of these multiplets. 
\end{appendices}
\vfill

\renewcommand\bibname{Bibliography}
\bibliography{references}{}

\providecommand{\href}[2]{#2}\begingroup\raggedright\begin{thebibliography}{10}

\bibitem{Froggatt:1978nt}
C.~D. Froggatt and H.~B. Nielsen, ``{Hierarchy of Quark Masses, Cabibbo Angles
  and {CP} Violation},''
  \href{http://dx.doi.org/10.1016/0550-3213(79)90316-X}{{\em Nucl. Phys. B}
  {\bfseries 147} (1979) 277--298}.

\bibitem{Maki:1962mu}
Z.~Maki, M.~Nakagawa, and S.~Sakata, ``{Remarks on the unified model of
  elementary particles},'' \href{http://dx.doi.org/10.1143/PTP.28.870}{{\em
  Prog. Theor. Phys.} {\bfseries 28} (1962) 870--880}.

\bibitem{Pontecorvo:1957qd}
B.~Pontecorvo, ``{Inverse beta processes and nonconservation of lepton
  charge},'' {\em Zh. Eksp. Teor. Fiz.} {\bfseries 34} (1957) 247.

\bibitem{Altarelli:2010gt}
G.~Altarelli and F.~Feruglio, ``{Discrete Flavor Symmetries and Models of
  Neutrino Mixing},'' \href{http://dx.doi.org/10.1103/RevModPhys.82.2701}{{\em
  Rev. Mod. Phys.} {\bfseries 82} (2010) 2701--2729},
  \href{http://arxiv.org/abs/1002.0211}{{\ttfamily 1002.0211 [hep-ph]}}.

\bibitem{Feruglio:2007hi}
F.~Feruglio and Y.~Lin, ``{Fermion Mass Hierarchies and Flavour Mixing from a
  Minimal Discrete Symmetry},''
  \href{http://dx.doi.org/10.1016/j.nuclphysb.2008.02.008}{{\em Nucl. Phys. B}
  {\bfseries 800} (2008) 77--93},
  \href{http://arxiv.org/abs/0712.1528}{{\ttfamily 0712.1528 [hep-ph]}}.

\bibitem{Altarelli:2005yx}
G.~Altarelli and F.~Feruglio, ``{Tri-bimaximal neutrino mixing, {$A_4$} and the
  modular symmetry},''
  \href{http://dx.doi.org/10.1016/j.nuclphysb.2006.02.015}{{\em Nucl. Phys. B}
  {\bfseries 741} (2006) 215--235},
  \href{http://arxiv.org/abs/hep-ph/0512103}{{\ttfamily hep-ph/0512103}}.

\bibitem{Lam:2008sh}
C.~S. Lam, ``{The Unique Horizontal Symmetry of Leptons},''
  \href{http://dx.doi.org/10.1103/PhysRevD.78.073015}{{\em Phys. Rev. D}
  {\bfseries 78} (2008) 073015},
  \href{http://arxiv.org/abs/0809.1185}{{\ttfamily 0809.1185 [hep-ph]}}.

\bibitem{Feruglio:2017spp}
F.~Feruglio, {\em {Are neutrino masses modular forms?}},
  \href{http://dx.doi.org/10.1142/9789813238053_0012}{pp.~227--266}.
\newblock 2019.
\newblock \href{http://arxiv.org/abs/1706.08749}{{\ttfamily 1706.08749
  [hep-ph]}}.

\bibitem{Lauer:1989ax}
J.~Lauer, J.~Mas, and H.~P. Nilles, ``{Duality and the Role of Nonperturbative
  Effects on the World Sheet},''
  \href{http://dx.doi.org/10.1016/0370-2693(89)91190-8}{{\em Phys. Lett. B}
  {\bfseries 226} (1989) 251--256}.

\bibitem{Lauer:1990tm}
J.~Lauer, J.~Mas, and H.~P. Nilles, ``{Twisted sector representations of
  discrete background symmetries for two-dimensional orbifolds},''
  \href{http://dx.doi.org/10.1016/0550-3213(91)90095-F}{{\em Nucl. Phys. B}
  {\bfseries 351} (1991) 353--424}.

\bibitem{Kobayashi:2018vbk}
T.~Kobayashi, K.~Tanaka, and T.~H. Tatsuishi, ``{Neutrino mixing from finite
  modular groups},'' \href{http://dx.doi.org/10.1103/PhysRevD.98.016004}{{\em
  Phys. Rev. D} {\bfseries 98} no.~1, (2018) 016004},
  \href{http://arxiv.org/abs/1803.10391}{{\ttfamily 1803.10391 [hep-ph]}}.

\bibitem{Kobayashi:2018wkl}
T.~Kobayashi, Y.~Shimizu, K.~Takagi, M.~Tanimoto, T.~H. Tatsuishi, and
  H.~Uchida, ``{Finite modular subgroups for fermion mass matrices and
  baryon/lepton number violation},''
  \href{http://dx.doi.org/10.1016/j.physletb.2019.05.034}{{\em Phys. Lett. B}
  {\bfseries 794} (2019) 114--121},
  \href{http://arxiv.org/abs/1812.11072}{{\ttfamily 1812.11072 [hep-ph]}}.

\bibitem{Criado:2018thu}
J.~C. Criado and F.~Feruglio, ``{Modular Invariance Faces Precision Neutrino
  Data},'' \href{http://dx.doi.org/10.21468/SciPostPhys.5.5.042}{{\em SciPost
  Phys.} {\bfseries 5} no.~5, (2018) 042},
  \href{http://arxiv.org/abs/1807.01125}{{\ttfamily 1807.01125 [hep-ph]}}.

\bibitem{Kobayashi:2018scp}
T.~Kobayashi, N.~Omoto, Y.~Shimizu, K.~Takagi, M.~Tanimoto, and T.~H.
  Tatsuishi, ``{Modular A$_{4}$ invariance and neutrino mixing},''
  \href{http://dx.doi.org/10.1007/JHEP11(2018)196}{{\em JHEP} {\bfseries 11}
  (2018) 196}, \href{http://arxiv.org/abs/1808.03012}{{\ttfamily 1808.03012
  [hep-ph]}}.

\bibitem{Okada:2018yrn}
H.~Okada and M.~Tanimoto, ``{CP violation of quarks in {$A_4$} modular
  invariance},'' \href{http://dx.doi.org/10.1016/j.physletb.2019.02.028}{{\em
  Phys. Lett. B} {\bfseries 791} (2019) 54--61},
  \href{http://arxiv.org/abs/1812.09677}{{\ttfamily 1812.09677 [hep-ph]}}.

\bibitem{Okada:2019uoy}
H.~Okada and M.~Tanimoto, ``{Towards unification of quark and lepton flavors in
  {$A_4$} modular invariance},''
  \href{http://dx.doi.org/10.1140/epjc/s10052-021-08845-y}{{\em Eur. Phys. J.
  C} {\bfseries 81} no.~1, (2021) 52},
  \href{http://arxiv.org/abs/1905.13421}{{\ttfamily 1905.13421 [hep-ph]}}.

\bibitem{Ding:2019zxk}
G.-J. Ding, S.~F. King, and X.-G. Liu, ``{Modular A$_{4}$ symmetry models of
  neutrinos and charged leptons},''
  \href{http://dx.doi.org/10.1007/JHEP09(2019)074}{{\em JHEP} {\bfseries 09}
  (2019) 074}, \href{http://arxiv.org/abs/1907.11714}{{\ttfamily 1907.11714
  [hep-ph]}}.

\bibitem{Kobayashi:2019xvz}
T.~Kobayashi, Y.~Shimizu, K.~Takagi, M.~Tanimoto, and T.~H. Tatsuishi,
  ``{{$A_4$} lepton flavor model and modulus stabilization from {$S_4$} modular
  symmetry},'' \href{http://dx.doi.org/10.1103/PhysRevD.100.115045}{{\em Phys.
  Rev. D} {\bfseries 100} no.~11, (2019) 115045},
  \href{http://arxiv.org/abs/1909.05139}{{\ttfamily 1909.05139 [hep-ph]}}.
  [Erratum: Phys.Rev.D 101, 039904 (2020)].

\bibitem{Asaka:2019vev}
T.~Asaka, Y.~Heo, T.~H. Tatsuishi, and T.~Yoshida, ``{Modular $A_4$ invariance
  and leptogenesis},'' \href{http://dx.doi.org/10.1007/JHEP01(2020)144}{{\em
  JHEP} {\bfseries 01} (2020) 144},
  \href{http://arxiv.org/abs/1909.06520}{{\ttfamily 1909.06520 [hep-ph]}}.

\bibitem{Ding:2019gof}
G.-J. Ding, S.~F. King, X.-G. Liu, and J.-N. Lu, ``{Modular S$_{4}$ and A$_{4}$
  symmetries and their fixed points: new predictive examples of lepton
  mixing},'' \href{http://dx.doi.org/10.1007/JHEP12(2019)030}{{\em JHEP}
  {\bfseries 12} (2019) 030}, \href{http://arxiv.org/abs/1910.03460}{{\ttfamily
  1910.03460 [hep-ph]}}.

\bibitem{Zhang:2019ngf}
D.~Zhang, ``{A modular $A_4$ symmetry realization of two-zero textures of the
  Majorana neutrino mass matrix},''
  \href{http://dx.doi.org/10.1016/j.nuclphysb.2020.114935}{{\em Nucl. Phys. B}
  {\bfseries 952} (2020) 114935},
  \href{http://arxiv.org/abs/1910.07869}{{\ttfamily 1910.07869 [hep-ph]}}.

\bibitem{King:2020qaj}
S.~J.~D. King and S.~F. King, ``{Fermion mass hierarchies from modular
  symmetry},'' \href{http://dx.doi.org/10.1007/JHEP09(2020)043}{{\em JHEP}
  {\bfseries 09} (2020) 043}, \href{http://arxiv.org/abs/2002.00969}{{\ttfamily
  2002.00969 [hep-ph]}}.

\bibitem{Ding:2020yen}
G.-J. Ding and F.~Feruglio, ``{Testing Moduli and Flavon Dynamics with Neutrino
  Oscillations},'' \href{http://dx.doi.org/10.1007/JHEP06(2020)134}{{\em JHEP}
  {\bfseries 06} (2020) 134}, \href{http://arxiv.org/abs/2003.13448}{{\ttfamily
  2003.13448 [hep-ph]}}.

\bibitem{Asaka:2020tmo}
T.~Asaka, Y.~Heo, and T.~Yoshida, ``{Lepton flavor model with modular $A_4$
  symmetry in large volume limit},''
  \href{http://dx.doi.org/10.1016/j.physletb.2020.135956}{{\em Phys. Lett. B}
  {\bfseries 811} (2020) 135956},
  \href{http://arxiv.org/abs/2009.12120}{{\ttfamily 2009.12120 [hep-ph]}}.

\bibitem{Okada:2020brs}
H.~Okada and M.~Tanimoto, ``{Spontaneous CP violation by modulus $\tau$ in
  $A_4$ model of lepton flavors},''
  \href{http://dx.doi.org/10.1007/JHEP03(2021)010}{{\em JHEP} {\bfseries 03}
  (2021) 010}, \href{http://arxiv.org/abs/2012.01688}{{\ttfamily 2012.01688
  [hep-ph]}}.

\bibitem{Yao:2020qyy}
C.-Y. Yao, J.-N. Lu, and G.-J. Ding, ``{Modular Invariant $A_{4}$ Models for
  Quarks and Leptons with Generalized CP Symmetry},''
  \href{http://dx.doi.org/10.1007/JHEP05(2021)102}{{\em JHEP} {\bfseries 05}
  (2021) 102}, \href{http://arxiv.org/abs/2012.13390}{{\ttfamily 2012.13390
  [hep-ph]}}.

\bibitem{Okada:2021qdf}
H.~Okada, Y.~Shimizu, M.~Tanimoto, and T.~Yoshida, ``{Modulus \ensuremath{\tau}
  linking leptonic CP violation to baryon asymmetry in A$_{4}$ modular
  invariant flavor model},''
  \href{http://dx.doi.org/10.1007/JHEP07(2021)184}{{\em JHEP} {\bfseries 07}
  (2021) 184}, \href{http://arxiv.org/abs/2105.14292}{{\ttfamily 2105.14292
  [hep-ph]}}.

\bibitem{Nomura:2021yjb}
T.~Nomura, H.~Okada, and Y.~Orikasa, ``{Quark and lepton flavor model with
  leptoquarks in a modular {$A_4$} symmetry},''
  \href{http://dx.doi.org/10.1140/epjc/s10052-021-09667-8}{{\em Eur. Phys. J.
  C} {\bfseries 81} no.~10, (2021) 947},
  \href{http://arxiv.org/abs/2106.12375}{{\ttfamily 2106.12375 [hep-ph]}}.

\bibitem{Chen:2021prl}
M.-C. Chen, V.~Knapp-Perez, M.~Ramos-Hamud, S.~Ramos-Sanchez, M.~Ratz, and
  S.~Shukla, ``{Quasi\textendash{}eclectic modular flavor symmetries},''
  \href{http://dx.doi.org/10.1016/j.physletb.2021.136843}{{\em Phys. Lett. B}
  {\bfseries 824} (2022) 136843},
  \href{http://arxiv.org/abs/2108.02240}{{\ttfamily 2108.02240 [hep-ph]}}.

\bibitem{Nomura:2022boj}
T.~Nomura, H.~Okada, and Y.~Shoji, ``{$SU(4)_C \times SU(2)_L \times U(1)_R$
  models with modular $A_4$ symmetry},''
  \href{http://arxiv.org/abs/2206.04466}{{\ttfamily 2206.04466 [hep-ph]}}.

\bibitem{Gunji:2022xig}
Y.~Gunji, K.~Ishiwata, and T.~Yoshida, ``{Subcritical regime of hybrid
  inflation with modular {$A_{4}$} symmetry},''
  \href{http://dx.doi.org/10.1007/JHEP11(2022)002}{{\em JHEP} {\bfseries 11}
  (2022) 002}, \href{http://arxiv.org/abs/2208.10086}{{\ttfamily 2208.10086
  [hep-ph]}}.

\bibitem{Penedo:2018nmg}
J.~T. Penedo and S.~T. Petcov, ``{Lepton Masses and Mixing from Modular {$S_4$}
  Symmetry},'' \href{http://dx.doi.org/10.1016/j.nuclphysb.2018.12.016}{{\em
  Nucl. Phys. B} {\bfseries 939} (2019) 292--307},
  \href{http://arxiv.org/abs/1806.11040}{{\ttfamily 1806.11040 [hep-ph]}}.

\bibitem{Novichkov:2018ovf}
P.~P. Novichkov, J.~T. Penedo, S.~T. Petcov, and A.~V. Titov, ``{Modular
  S$_{4}$ models of lepton masses and mixing},''
  \href{http://dx.doi.org/10.1007/JHEP04(2019)005}{{\em JHEP} {\bfseries 04}
  (2019) 005}, \href{http://arxiv.org/abs/1811.04933}{{\ttfamily 1811.04933
  [hep-ph]}}.

\bibitem{deMedeirosVarzielas:2019cyj}
I.~de~Medeiros~Varzielas, S.~F. King, and Y.-L. Zhou, ``{Multiple modular
  symmetries as the origin of flavor},''
  \href{http://dx.doi.org/10.1103/PhysRevD.101.055033}{{\em Phys. Rev. D}
  {\bfseries 101} no.~5, (2020) 055033},
  \href{http://arxiv.org/abs/1906.02208}{{\ttfamily 1906.02208 [hep-ph]}}.

\bibitem{Kobayashi:2019mna}
T.~Kobayashi, Y.~Shimizu, K.~Takagi, M.~Tanimoto, and T.~H. Tatsuishi, ``{New
  {$A_4$} lepton flavor model from {$S_4$} modular symmetry},''
  \href{http://dx.doi.org/10.1007/JHEP02(2020)097}{{\em JHEP} {\bfseries 02}
  (2020) 097}, \href{http://arxiv.org/abs/1907.09141}{{\ttfamily 1907.09141
  [hep-ph]}}.

\bibitem{King:2019vhv}
S.~F. King and Y.-L. Zhou, ``{Trimaximal {TM$_1$} mixing with two modular
  {$S_4$} groups},'' \href{http://dx.doi.org/10.1103/PhysRevD.101.015001}{{\em
  Phys. Rev. D} {\bfseries 101} no.~1, (2020) 015001},
  \href{http://arxiv.org/abs/1908.02770}{{\ttfamily 1908.02770 [hep-ph]}}.

\bibitem{Criado:2019tzk}
J.~C. Criado, F.~Feruglio, and S.~J.~D. King, ``{Modular Invariant Models of
  Lepton Masses at Levels 4 and 5},''
  \href{http://dx.doi.org/10.1007/JHEP02(2020)001}{{\em JHEP} {\bfseries 02}
  (2020) 001}, \href{http://arxiv.org/abs/1908.11867}{{\ttfamily 1908.11867
  [hep-ph]}}.

\bibitem{Wang:2019ovr}
X.~Wang and S.~Zhou, ``{The minimal seesaw model with a modular S$_{4}$
  symmetry},'' \href{http://dx.doi.org/10.1007/JHEP05(2020)017}{{\em JHEP}
  {\bfseries 05} (2020) 017}, \href{http://arxiv.org/abs/1910.09473}{{\ttfamily
  1910.09473 [hep-ph]}}.

\bibitem{Wang:2020dbp}
X.~Wang, ``{Dirac neutrino mass models with a modular {$S_4$} symmetry},''
  \href{http://dx.doi.org/10.1016/j.nuclphysb.2020.115247}{{\em Nucl. Phys. B}
  {\bfseries 962} (2021) 115247},
  \href{http://arxiv.org/abs/2007.05913}{{\ttfamily 2007.05913 [hep-ph]}}.

\bibitem{Qu:2021jdy}
B.-Y. Qu, X.-G. Liu, P.-T. Chen, and G.-J. Ding, ``{Flavor mixing and CP
  violation from the interplay of an {$S_4$} modular group and a generalized CP
  symmetry},'' \href{http://dx.doi.org/10.1103/PhysRevD.104.076001}{{\em Phys.
  Rev. D} {\bfseries 104} no.~7, (2021) 076001},
  \href{http://arxiv.org/abs/2106.11659}{{\ttfamily 2106.11659 [hep-ph]}}.

\bibitem{Novichkov:2018nkm}
P.~P. Novichkov, J.~T. Penedo, S.~T. Petcov, and A.~V. Titov, ``{Modular
  {$A_{5}$} symmetry for flavour model building},''
  \href{http://dx.doi.org/10.1007/JHEP04(2019)174}{{\em JHEP} {\bfseries 04}
  (2019) 174}, \href{http://arxiv.org/abs/1812.02158}{{\ttfamily 1812.02158
  [hep-ph]}}.

\bibitem{Ding:2019xna}
G.-J. Ding, S.~F. King, and X.-G. Liu, ``{Neutrino mass and mixing with {$A_5$}
  modular symmetry},''
  \href{http://dx.doi.org/10.1103/PhysRevD.100.115005}{{\em Phys. Rev. D}
  {\bfseries 100} no.~11, (2019) 115005},
  \href{http://arxiv.org/abs/1903.12588}{{\ttfamily 1903.12588 [hep-ph]}}.

\bibitem{Feruglio:2019ybq}
F.~Feruglio and A.~Romanino, ``{Lepton flavor symmetries},''
  \href{http://dx.doi.org/10.1103/RevModPhys.93.015007}{{\em Rev. Mod. Phys.}
  {\bfseries 93} no.~1, (2021) 015007},
  \href{http://arxiv.org/abs/1912.06028}{{\ttfamily 1912.06028 [hep-ph]}}.

\bibitem{Novichkov:2019sqv}
P.~P. Novichkov, J.~T. Penedo, S.~T. Petcov, and A.~V. Titov, ``{Generalised
  {CP} Symmetry in Modular-Invariant Models of Flavour},''
  \href{http://dx.doi.org/10.1007/JHEP07(2019)165}{{\em JHEP} {\bfseries 07}
  (2019) 165}, \href{http://arxiv.org/abs/1905.11970}{{\ttfamily 1905.11970
  [hep-ph]}}.

\bibitem{Baur:2019kwi}
A.~Baur, H.~P. Nilles, A.~Trautner, and P.~K.~S. Vaudrevange, ``{Unification of
  Flavor, CP, and Modular Symmetries},''
  \href{http://dx.doi.org/10.1016/j.physletb.2019.03.066}{{\em Phys. Lett. B}
  {\bfseries 795} (2019) 7--14},
  \href{http://arxiv.org/abs/1901.03251}{{\ttfamily 1901.03251 [hep-th]}}.

\bibitem{Acharya:1995ag}
B.~S. Acharya, D.~Bailin, A.~Love, W.~A. Sabra, and S.~Thomas, ``{Spontaneous
  breaking of CP symmetry by orbifold moduli},''
  \href{http://dx.doi.org/10.1016/0370-2693(95)00945-H}{{\em Phys. Lett. B}
  {\bfseries 357} (1995) 387--396},
  \href{http://arxiv.org/abs/hep-th/9506143}{{\ttfamily hep-th/9506143}}.
  [Erratum: Phys.Lett.B 407, 451--451 (1997)].

\bibitem{Dent:2001cc}
T.~Dent, ``{CP violation and modular symmetries},''
  \href{http://dx.doi.org/10.1103/PhysRevD.64.056005}{{\em Phys. Rev. D}
  {\bfseries 64} (2001) 056005},
  \href{http://arxiv.org/abs/hep-ph/0105285}{{\ttfamily hep-ph/0105285}}.

\bibitem{Giedt:2002ns}
J.~Giedt, ``{CP violation and moduli stabilization in heterotic models},''
  \href{http://dx.doi.org/10.1142/S0217732302007879}{{\em Mod. Phys. Lett. A}
  {\bfseries 17} (2002) 1465--1473},
  \href{http://arxiv.org/abs/hep-ph/0204017}{{\ttfamily hep-ph/0204017}}.

\bibitem{Novichkov:2021evw}
P.~P. Novichkov, J.~T. Penedo, and S.~T. Petcov, ``{Fermion mass hierarchies,
  large lepton mixing and residual modular symmetries},''
  \href{http://dx.doi.org/10.1007/JHEP04(2021)206}{{\em JHEP} {\bfseries 04}
  (2021) 206}, \href{http://arxiv.org/abs/2102.07488}{{\ttfamily 2102.07488
  [hep-ph]}}.

\bibitem{Benes:2022bbg}
P.~Bene\v{s}, H.~Okada, and Y.~Orikasa, ``{Towards unification of lepton and
  quark mass matrices from double covering of modular $A_4$ flavor symmetry},''
  \href{http://arxiv.org/abs/2212.07245}{{\ttfamily 2212.07245 [hep-ph]}}.

\bibitem{Chen:2019ewa}
M.-C. Chen, S.~Ramos-S\'anchez, and M.~Ratz, ``{A note on the predictions of
  models with modular flavor symmetries},''
  \href{http://dx.doi.org/10.1016/j.physletb.2019.135153}{{\em Phys. Lett. B}
  {\bfseries 801} (2020) 135153},
  \href{http://arxiv.org/abs/1909.06910}{{\ttfamily 1909.06910 [hep-ph]}}.

\bibitem{Du:2020ylx}
X.~Du and F.~Wang, ``{SUSY breaking constraints on modular flavor $S_{3}$
  invariant SU(5) GUT model},''
  \href{http://dx.doi.org/10.1007/JHEP02(2021)221}{{\em JHEP} {\bfseries 02}
  (2021) 221}, \href{http://arxiv.org/abs/2012.01397}{{\ttfamily 2012.01397
  [hep-ph]}}.

\bibitem{Kobayashi:2019rzp}
T.~Kobayashi, Y.~Shimizu, K.~Takagi, M.~Tanimoto, and T.~H. Tatsuishi,
  ``{Modular $S_3$-invariant flavor model in SU(5) grand unified theory},''
  \href{http://dx.doi.org/10.1093/ptep/ptaa055}{{\em PTEP} {\bfseries 2020}
  no.~5, (2020) 053B05}, \href{http://arxiv.org/abs/1906.10341}{{\ttfamily
  1906.10341 [hep-ph]}}.

\bibitem{Okada:2019xqk}
H.~Okada and Y.~Orikasa, ``{Modular {$S_3$} symmetric radiative seesaw
  model},'' \href{http://dx.doi.org/10.1103/PhysRevD.100.115037}{{\em Phys.
  Rev. D} {\bfseries 100} no.~11, (2019) 115037},
  \href{http://arxiv.org/abs/1907.04716}{{\ttfamily 1907.04716 [hep-ph]}}.

\bibitem{Mishra:2020gxg}
S.~Mishra, ``{Neutrino mixing and Leptogenesis with modular {$S_3$} symmetry in
  the framework of type III seesaw},''
  \href{http://arxiv.org/abs/2008.02095}{{\ttfamily 2008.02095 [hep-ph]}}.

\bibitem{Capozzi:2021fjo}
F.~Capozzi, E.~Di~Valentino, E.~Lisi, A.~Marrone, A.~Melchiorri, and
  A.~Palazzo, ``{Unfinished fabric of the three neutrino paradigm},''
  \href{http://dx.doi.org/10.1103/PhysRevD.104.083031}{{\em Phys. Rev. D}
  {\bfseries 104} no.~8, (2021) 083031},
  \href{http://arxiv.org/abs/2107.00532}{{\ttfamily 2107.00532 [hep-ph]}}.

\bibitem{eBOSS:2020yzd}
{\bfseries eBOSS} Collaboration, S.~Alam {\em et~al.}, ``{Completed SDSS-IV
  extended Baryon Oscillation Spectroscopic Survey: Cosmological implications
  from two decades of spectroscopic surveys at the Apache Point Observatory},''
  \href{http://dx.doi.org/10.1103/PhysRevD.103.083533}{{\em Phys. Rev. D}
  {\bfseries 103} no.~8, (2021) 083533},
  \href{http://arxiv.org/abs/2007.08991}{{\ttfamily 2007.08991 [astro-ph.CO]}}.

\bibitem{KamLAND-Zen:2022tow}
{\bfseries KamLAND-Zen} Collaboration, S.~Abe {\em et~al.}, ``{Search for the
  Majorana Nature of Neutrinos in the Inverted Mass Ordering Region with
  KamLAND-Zen},'' \href{http://dx.doi.org/10.1103/PhysRevLett.130.051801}{{\em
  Phys. Rev. Lett.} {\bfseries 130} no.~5, (2023) 051801},
  \href{http://arxiv.org/abs/2203.02139}{{\ttfamily 2203.02139 [hep-ex]}}.

\bibitem{KATRIN:2021uub}
{\bfseries KATRIN} Collaboration, M.~Aker {\em et~al.}, ``{Direct neutrino-mass
  measurement with sub-electronvolt sensitivity},''
  \href{http://dx.doi.org/10.1038/s41567-021-01463-1}{{\em Nature Phys.}
  {\bfseries 18} no.~2, (2022) 160--166},
  \href{http://arxiv.org/abs/2105.08533}{{\ttfamily 2105.08533 [hep-ex]}}.

\bibitem{Gunning1962}
R.~C. Gunning, \href{http://dx.doi.org/10.1515/9781400881666}{{\em Lectures on
  Modular Forms. ({AM}-48)}}.
\newblock Princeton University Press, Dec., 1962.

\bibitem{cohen:hal-01677348}
H.~Cohen and F.~Str{\"o}mberg, {\em {Modular Forms: A Classical Approach}},
  vol.~179 of {\em Graduate Studies in Mathematics}.
\newblock {American Mathematical Society}, 2017.

\bibitem{Liu:2019khw}
X.-G. Liu and G.-J. Ding, ``{Neutrino Masses and Mixing from Double Covering of
  Finite Modular Groups},''
  \href{http://dx.doi.org/10.1007/JHEP08(2019)134}{{\em JHEP} {\bfseries 08}
  (2019) 134}, \href{http://arxiv.org/abs/1907.01488}{{\ttfamily 1907.01488
  [hep-ph]}}.

\bibitem{Liu:2020akv}
X.-G. Liu, C.-Y. Yao, and G.-J. Ding, ``{Modular invariant quark and lepton
  models in double covering of {$S_4$} modular group},''
  \href{http://dx.doi.org/10.1103/PhysRevD.103.056013}{{\em Phys. Rev. D}
  {\bfseries 103} no.~5, (2021) 056013},
  \href{http://arxiv.org/abs/2006.10722}{{\ttfamily 2006.10722 [hep-ph]}}.

\bibitem{Wang:2020lxk}
X.~Wang, B.~Yu, and S.~Zhou, ``{Double covering of the modular {$A_5$} group
  and lepton flavor mixing in the minimal seesaw model},''
  \href{http://dx.doi.org/10.1103/PhysRevD.103.076005}{{\em Phys. Rev. D}
  {\bfseries 103} no.~7, (2021) 076005},
  \href{http://arxiv.org/abs/2010.10159}{{\ttfamily 2010.10159 [hep-ph]}}.

\bibitem{Yao:2020zml}
C.-Y. Yao, X.-G. Liu, and G.-J. Ding, ``{Fermion masses and mixing from the
  double cover and metaplectic cover of the {$A_5$} modular group},''
  \href{http://dx.doi.org/10.1103/PhysRevD.103.095013}{{\em Phys. Rev. D}
  {\bfseries 103} no.~9, (2021) 095013},
  \href{http://arxiv.org/abs/2011.03501}{{\ttfamily 2011.03501 [hep-ph]}}.

\bibitem{Novichkov:2020eep}
P.~P. Novichkov, J.~T. Penedo, and S.~T. Petcov, ``{Double cover of modular
  {$S_4$} for flavour model building},''
  \href{http://dx.doi.org/10.1016/j.nuclphysb.2020.115301}{{\em Nucl. Phys. B}
  {\bfseries 963} (2021) 115301},
  \href{http://arxiv.org/abs/2006.03058}{{\ttfamily 2006.03058 [hep-ph]}}.

\bibitem{deMedeirosVarzielas:2022fbw}
I.~de~Medeiros~Varzielas, S.~F. King, and M.~Levy, ``{Littlest modular
  seesaw},'' \href{http://dx.doi.org/10.1007/JHEP02(2023)143}{{\em JHEP}
  {\bfseries 02} (2023) 143}, \href{http://arxiv.org/abs/2211.00654}{{\ttfamily
  2211.00654 [hep-ph]}}.

\bibitem{Ding:2022nzn}
G.-J. Ding, X.-G. Liu, and C.-Y. Yao, ``{A minimal modular invariant neutrino
  model},'' \href{http://dx.doi.org/10.1007/JHEP01(2023)125}{{\em JHEP}
  {\bfseries 01} (2023) 125}, \href{http://arxiv.org/abs/2211.04546}{{\ttfamily
  2211.04546 [hep-ph]}}.

\bibitem{Novichkov:2022wvg}
P.~P. Novichkov, J.~T. Penedo, and S.~T. Petcov, ``{Modular flavour symmetries
  and modulus stabilisation},''
  \href{http://dx.doi.org/10.1007/JHEP03(2022)149}{{\em JHEP} {\bfseries 03}
  (2022) 149}, \href{http://arxiv.org/abs/2201.02020}{{\ttfamily 2201.02020
  [hep-ph]}}.

\bibitem{Knapp-Perez:2023nty}
V.~Knapp-Perez, X.-G. Liu, H.~P. Nilles, S.~Ramos-Sanchez, and M.~Ratz,
  ``{Matter matters in moduli fixing and modular flavor symmetries},''
  \href{http://arxiv.org/abs/2304.14437}{{\ttfamily 2304.14437 [hep-th]}}.

\bibitem{Ishiguro:2020tmo}
K.~Ishiguro, T.~Kobayashi, and H.~Otsuka, ``{Landscape of Modular Symmetric
  Flavor Models},'' \href{http://dx.doi.org/10.1007/JHEP03(2021)161}{{\em JHEP}
  {\bfseries 03} (2021) 161}, \href{http://arxiv.org/abs/2011.09154}{{\ttfamily
  2011.09154 [hep-ph]}}.

\bibitem{Ishiguro:2022pde}
K.~Ishiguro, H.~Okada, and H.~Otsuka, ``{Residual flavor symmetry breaking in
  the landscape of modular flavor models},''
  \href{http://dx.doi.org/10.1007/JHEP09(2022)072}{{\em JHEP} {\bfseries 09}
  (2022) 072}, \href{http://arxiv.org/abs/2206.04313}{{\ttfamily 2206.04313
  [hep-ph]}}.

\bibitem{Liu:2021gwa}
X.-G. Liu and G.-J. Ding, ``{Modular flavor symmetry and vector-valued modular
  forms},'' \href{http://dx.doi.org/10.1007/JHEP03(2022)123}{{\em JHEP}
  {\bfseries 03} (2022) 123}, \href{http://arxiv.org/abs/2112.14761}{{\ttfamily
  2112.14761 [hep-ph]}}.

\bibitem{Weinberg:1979sa}
S.~Weinberg, ``{Baryon and Lepton Nonconserving Processes},''
  \href{http://dx.doi.org/10.1103/PhysRevLett.43.1566}{{\em Phys. Rev. Lett.}
  {\bfseries 43} (1979) 1566--1570}.

\bibitem{Kuranaga:2021ujd}
H.~Kuranaga, H.~Ohki, and S.~Uemura, ``{Modular origin of mass hierarchy:
  Froggatt-Nielsen like mechanism},''
  \href{http://dx.doi.org/10.1007/JHEP07(2021)068}{{\em JHEP} {\bfseries 07}
  (2021) 068}, \href{http://arxiv.org/abs/2105.06237}{{\ttfamily 2105.06237
  [hep-ph]}}.

\bibitem{Feruglio:2021dte}
F.~Feruglio, V.~Gherardi, A.~Romanino, and A.~Titov, ``{Modular invariant
  dynamics and fermion mass hierarchies around $\tau = i$},''
  \href{http://dx.doi.org/10.1007/JHEP05(2021)242}{{\em JHEP} {\bfseries 05}
  (2021) 242}, \href{http://arxiv.org/abs/2101.08718}{{\ttfamily 2101.08718
  [hep-ph]}}.

\bibitem{Feruglio:2023bav}
F.~Feruglio, ``{Universal Predictions of Modular Invariant Flavor Models near
  the Self-Dual Point},''
  \href{http://dx.doi.org/10.1103/PhysRevLett.130.101801}{{\em Phys. Rev.
  Lett.} {\bfseries 130} no.~10, (2023) 101801}.

\bibitem{Novichkov:2021cgl}
P.~Novichkov, {\em {Aspects of the Modular Symmetry Approach to Lepton
  Flavour}}.
\newblock PhD thesis, SISSA, Trieste, 2021.

\bibitem{Antusch:2005gp}
S.~Antusch, J.~Kersten, M.~Lindner, M.~Ratz, and M.~A. Schmidt, ``{Running
  neutrino mass parameters in see-saw scenarios},''
  \href{http://dx.doi.org/10.1088/1126-6708/2005/03/024}{{\em JHEP} {\bfseries
  03} (2005) 024}, \href{http://arxiv.org/abs/hep-ph/0501272}{{\ttfamily
  hep-ph/0501272}}.

\bibitem{Altarelli:2010at}
G.~Altarelli and G.~Blankenburg, ``{Different $SO(10)$ Paths to Fermion Masses
  and Mixings},'' \href{http://dx.doi.org/10.1007/JHEP03(2011)133}{{\em JHEP}
  {\bfseries 03} (2011) 133}, \href{http://arxiv.org/abs/1012.2697}{{\ttfamily
  1012.2697 [hep-ph]}}.

\bibitem{Strait2016}
J.~Strait, , E.~McCluskey, T.~Lundin, J.~Willhite, T.~Hamernik,
  V.~Papadimitriou, A.~Marchionni, M.~J. Kim, M.~Nessi, D.~Montanari, and
  A.~Heavey, \href{http://dx.doi.org/10.2172/1250880}{``Long-baseline neutrino
  facility ({LBNF}) and deep underground neutrino experiment ({DUNE}):
  Conceptual design report. volume 3: Long-baseline neutrino facility for
  {DUNE},''} tech. rep., Jan., 2016.

\bibitem{Hyper-KamiokandeProto-:2015xww}
{\bfseries Hyper-Kamiokande Proto-} Collaboration, K.~Abe {\em et~al.},
  ``{Physics potential of a long-baseline neutrino oscillation experiment using
  a J-PARC neutrino beam and Hyper-Kamiokande},''
  \href{http://dx.doi.org/10.1093/ptep/ptv061}{{\em PTEP} {\bfseries 2015}
  (2015) 053C02}, \href{http://arxiv.org/abs/1502.05199}{{\ttfamily 1502.05199
  [hep-ex]}}.

\bibitem{nEXO:2018ylp}
{\bfseries nEXO} Collaboration, S.~A. Kharusi {\em et~al.}, ``{nEXO
  Pre-Conceptual Design Report},''
  \href{http://arxiv.org/abs/1805.11142}{{\ttfamily 1805.11142
  [physics.ins-det]}}.

\bibitem{Antusch:2003kp}
S.~Antusch, J.~Kersten, M.~Lindner, and M.~Ratz, ``{Running neutrino masses,
  mixings and CP phases: Analytical results and phenomenological
  consequences},''
  \href{http://dx.doi.org/10.1016/j.nuclphysb.2003.09.050}{{\em Nucl. Phys. B}
  {\bfseries 674} (2003) 401--433},
  \href{http://arxiv.org/abs/hep-ph/0305273}{{\ttfamily hep-ph/0305273}}.

\bibitem{Ereditato:2018eqf}
A.~Ereditato, ed., \href{http://dx.doi.org/10.1142/10600}{{\em {The State of
  the Art of Neutrino Physics}}}.
\newblock World Scientific, 2018.

\bibitem{Ishimori:2012zz}
H.~Ishimori, T.~Kobayashi, H.~Ohki, H.~Okada, Y.~Shimizu, and M.~Tanimoto,
  \href{http://dx.doi.org/10.1007/978-3-642-30805-5}{{\em {An introduction to
  non-Abelian discrete symmetries for particle physicists}}}, vol.~858.
\newblock 2012.

\end{thebibliography}\endgroup
\bibliographystyle{utphys}

\end{document}